\documentclass[pra,twocolumn,amsmath,amssymb,aps]{revtex4-2}

\usepackage{graphicx}
\usepackage{dcolumn}
\usepackage{bm}
\usepackage[utf8]{inputenc}
\usepackage{color}
\usepackage{hyperref}
\usepackage{harpoon}

\begin{document}

\title{Heisenberg-limited spin squeezing in a hybrid system with Silicon-Vacancy centers}

\author{Zhen-Qiang Ren}
\affiliation{School of Physics, Sun Yat-sen University, Guangzhou 510275, China}
\author{Xian-Liang Lu}
\affiliation{School of Physics, Sun Yat-sen University, Guangzhou 510275, China}
\author{Ze-Liang Xiang}
\email{xiangzliang@mail.sysu.edu.cn}
\affiliation{School of Physics, Sun Yat-sen University, Guangzhou 510275, China}

\date{\today}

\begin{abstract}
In this paper, we investigate spin squeezing in a hybrid quantum system consisting of a Silicon-Vacancy (SiV) center ensemble coupled to a diamond acoustic waveguide via the strain interaction. Two sets of non-overlapping driving fields, each contains two time-dependent  microwave fields, are applied to this hybrid system. By modulating these fields, the one-axis twist (OAT) interaction and two-axis two-spin (TATS) interaction can be independently realized. In the latter case the squeezing parameter scales to spin number as $\xi_R^2\sim1.61N^{-0.64}$ with the consideration of dissipation, which is very close to the Heisenberg limit. Furthermore, this hybrid system allows for the study of spin squeezing generated by the simultaneous presence of OAT and TATS interactions, which reveals sensitivity to the parity of the number of spins $N_{tot}$, whether it is even or odd. Our scheme enriches the approach for generating Heisenberg-limited spin squeezing in spin-phonon hybrid systems and offers the possibility for future applications in quantum information processing.

\end{abstract}

\maketitle

\section{Introduction}

The spin-squeezed state~\cite{Kitagawa1993,Wineland1994,Ma2011}, one of the typical quantum many-body entangled states~\cite{Zoller2001,Anders2001}, is a critical resource in quantum information processing~\cite{Cirac1995,Duan2001,Kimble2008,Ma2018,Khatri2021} and quantum metrology~\cite{Giovannetti2006,Max2010,Wang2018}. Such a state can be used to overcome the shot-noise limit~\cite{Wolfgramm2010,Onur2016,Pezz2018,Chabuda2020} and to study many-body entanglement~\cite{Matteo2018,Tura2014,Han2020,Hans2009,Korbicz2005}. The ability to efficiently generate spin-squeezed states is the first step in unlocking its potential applications. There are various approaches to produce spin squeezing, such as transferring the squeezing from squeezed light to spin ensembles~\cite{Kuzmich1997,Hald1999,Vernac2000,Fleischhauer2002}, quantum nondemolition (QND) measurements of collective spins~\cite{Kuzmich1999,Kuzmich2000,Inoue2013,Rossi2020}, and utilizing the OAT and two-axis twist (TAT) squeezing interactions~\cite{Kitagawa1993,Ma2011,Liu2011,Zhang2017,Groszkowski2020,Huang2021,Zhang2014,Huang2015,Bai2021}. While the OAT interaction is able to deterministically generate spin squeezing and has been effectively implemented in different experimental systems, including the atomic systems~\cite{Helmerson2001,Shenchao2021,Zhang2003} and trapped ion systems~\cite{Justin2016}, it cannot reach the Heisenberg limit in principle~\cite{Ma2011}. Conversely, although the TAT interaction allows for Heisenberg-limited spin squeezing, it presents various implementation challenges. Thus, many efforts have been devoted to engineering the TAT interaction and generating Heisenberg-limited spin squeezing in current state-of-the-art systems ~\cite{Ma2011,Groszkowski2020,Liu2011,Wu2015,Wang2017,Kitzinger2020,Witkowska2022,Huang2021,Huang2023}.

The quantum hybrid systems based on color centers have been extensively developed in quantum information processing benefiting from their longer coherence time and better controllability~\cite{Sipahigil2014,Li2016,Golter2016,Song2017,Lipengbo2020,Zhou2022,Zhou2021,Bennett2013,Xia2016,MA2016,Li2020,Chen2021}. The SiV center, a type of color centers, has the fourfold-degenerate ground state, which enables various interactions between corresponding electronic levels~\cite{Hepp2014,Sipahigil2016,Jingyuan2017,Meesala2018,Lemonde2018}. Moreover, due to the strong coupling between SiV centers and acoustic modes, the entanglement of SiV centers can be generated mediated by phonons~\cite{Lemonde2018,Qiao2020,Ren2022}. Such a strong coupling can also be employed to achieve better spin squeezing in spin-phonon hybrid systems based on SiV centers~\cite{Kepesidis2016,Sohn2018,Lemonde2018,Qiao2020,Dong2021,Zhou2022}.

In this work, we propose a scheme for generating spin-squeezed states in a hybrid system consisiting of an ensemble of SiV centers coupled to the acoustic mode of a diamond waveguide via the strain interaction. This SiV ensemble is partitioned into two different segments resulting from two sets of non-overlapping microwave fields. The strain-induced coupling enables effective spin-spin interactions mediated by virtual phonons, then the OAT and TATS interactions can be induced independently, where the latter one can realize Heisenberg-limited spin squeezing~\cite{Kitagawa1993,Wineland1994,Ma2011}. Furthermore, we investigate the spin-squeezed states generated by the mixed Hamiltonian of OAT and TATS interactions and show the sensitivity of these states to the even-odd spin particles, which holds potential for sensing applications. Considering practical dissipations in the system, the squeezing parameter $\xi_R^2$ has a trend as $\xi_R^2\sim1.61N^{-0.64}$, which can be used to achieve a measurement precision close to the Heisenberg limit. Compared to other schemes that necessitate the use of squeezed field injection, complex pulse drive or parametric drive to generate better spin-squeezed states, our scheme requires only the appropriate modulation of microwave fields and allows better spin-squeezed states based on this spin-phonon hybrid system.

Our paper is organized as follows. In Sec.~II, we introduce the theoretical model of a hybrid quantum system consisting of two SiV-center segments embedded in a quasi 1D acoustic waveguide. Section III shows the time evolution of squeezing parameters $ \xi_S^2 $ and $ \xi_R^2 $ in the case of the OAT, TATS and mixed OAT-TATS Hamiltonians. In Sec.~IV, we discuss the experimental feasibility of this scheme and analyze the influence caused by the experimental dissipation of this hybrid system. Finally, we make a summary in Sec.~V.

\section{Model}

%=================================================================== 
\begin{figure}[tpb]
\centering
\includegraphics[width=1\linewidth]{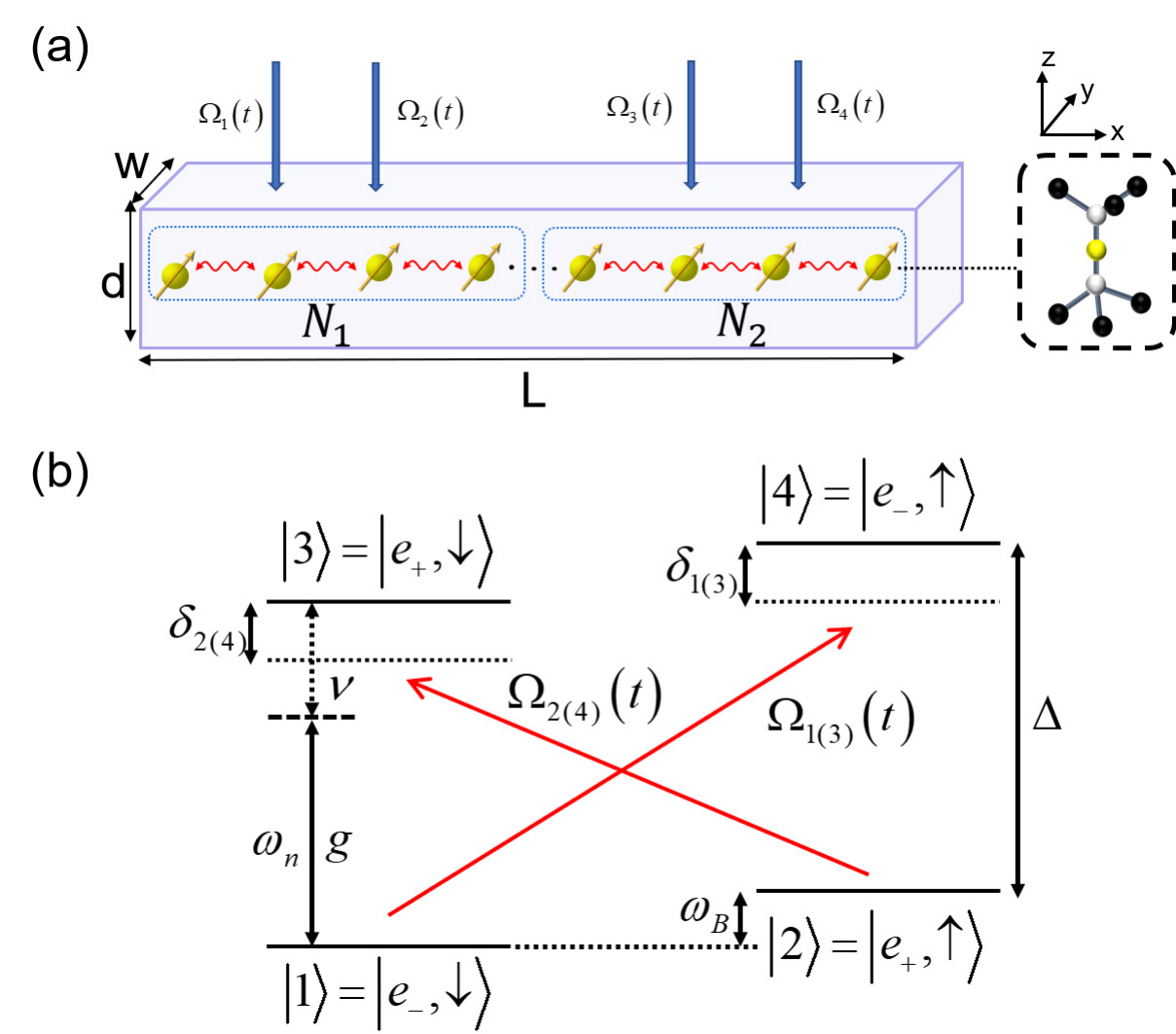}
\caption{(a) Sketch of an array of SiV centers embedded in a 1D diamond waveguide. The length, width, and thickness of the waveguide are L, w, d, respectively. Molecular structure of the SiV center is shown as the inset. In this system, there are two different segments $S_1$ and $S_2$ in the SiV-center ensemble, which contains $N_1$ and $N_2$ SiV centers, respectively, resulted from the different set of driving fields. (b) The level structure of the electronic ground state of the SiV center. The time-dependent microwave driving fields induce the transitions between levels $\vert 1\rangle\leftrightarrow\vert 4\rangle$ and $\vert 2\rangle\leftrightarrow\vert 3\rangle$, while the transitions between levels $\vert 1\rangle\leftrightarrow\vert 3\rangle$ and $\vert 2\rangle\leftrightarrow\vert 4\rangle$ are caused by the strain-induced coupling.}
\label{fig1}
\end{figure}
%=================================================================== 

As depicted in Fig.~\ref{fig1} (a), we consider a spin-phonon hybrid system, where two sets of $N_1$ and $N_2$ SiV centers in segment $S_1$ and segment $S_2$, respectively, are coupled to an acoustic mode of a 1D diamond waveguide via the strain-induced interaction. This interaction arises from the change of Coulomb energy of the electronic states due to the displacement of atoms forming the defect. First, we consider the SiV centers in segment $S_1$ which are driven by two time-dependent microwave fields $\Omega_1(t)$ and $\Omega_2(t)$, and this system can be described by the Hamiltonian~\cite{Lemonde2018,Ren2022}
\begin{equation}\label{eq:one}
H_{S_1}=H_{\rm SiV_{S_1}}+H_{\rm ph}+H_{\rm strain_{S_1}},
\end{equation}
where $H_{\rm SiV_{S_1}}$ and $H_{\rm ph}$ are the Hamiltonians of SiV centers in segment $S_1$ and the acoustic mode, respectively, and $H_{\rm strain_{S_1}}$ denotes the strain-induced coupling between the orbital degree of the SiV center in segment $S_1$ and the common acoustic mode of the waveguide, as shown in Fig.~\ref{fig1}(b).

The SiV center is an interstitial point defect in which a silicon atom is positioned midway between two adjacent missing carbon atoms in the diamond lattice, as depicted in the inset of Fig.~\ref{fig1}(a). Its ground state is four-fold degenerate, with the corresponding energy splitting $\Delta=[\lambda_g^2+\left(\Upsilon_x^2+\Upsilon_y^2\right)]^{1/2}\approx2\pi\times46$~GHz, where $\lambda_g=2\pi\times45$~GHz is the spin-orbit coupling strength, and $\Upsilon_{x(y)}$ describes the strength of the Jahn-Teller (JT) effect along $\vec x(\vec y)$ direction~\cite{Hepp2014,Lemonde2018}. Two time-dependent microwave fields $\Omega_{1,2}(t)$ with frequencies $\omega_{1,2}$ will induce transitions between states $\vert 1\rangle\leftrightarrow\vert 4\rangle$ and $\vert 2\rangle\leftrightarrow\vert  3\rangle$, as shown in Fig.~\ref{fig1}(b). Consequently, the dynamics of SiV centers can be described by the Hamiltonian$ (\hbar=1) $~\cite{Lemonde2018,Ren2022} 
\begin{equation}
\begin{split}
H_{\rm SiV_{S_1}}=&\sum_j^{N_1}[\omega_B\vert 2\rangle_j \langle 2\vert +\Delta\vert 3\rangle_j \langle 3\vert +(\omega_B+\Delta)\vert 4\rangle_j \langle 4\vert\\
&+\frac{\Omega_1(t)}{2}\vert 1\rangle_j \langle 4\vert e^{i\omega_1t}
+\frac{\Omega_2(t)}{2}\vert 2\rangle_j \langle 3\vert e^{i\omega_2t} 
\left.+{\rm H.c.}\right],
\end{split}
\label{eq:two}
\end{equation}
where $\omega_B=\gamma_sB_0$ denotes the energy-level splitting induced by the Zeeman effect, and we set $\omega_B\approx2\pi\times5$~GHz here. $\gamma_s$ is the spin gyromagnetic ratio, and $j$ labels the $j$-th SiV center in segment $S_1$.

Now we consider acoustic modes in the quasi 1D diamond waveguide. The length, width, and thickness of the waveguide are $L$, $w$, $d$, respectively, as shown in Fig.~\ref{fig1}(a), satisfying $L\gg w,d $ meanwhile. The quantized Hamiltonian of acoustic modes can be written as 
\begin{equation}
H_{\rm ph}=\sum_{n,k}\omega_{n,k}a_{n,k}^{\dag}a_{n,k},
\label{eq:five}
\end{equation}
where $a_{n,k}$ is the annihilation operator of one acoustic mode.

Within the framework of linear elasticity theory, the strain-induced coupling Hamiltonian between SiV centers in segment $S_1$ and acoustic modes in the waveguide is given by~\cite{Lemonde2018,Hepp2014}
\begin{equation}
H_{\rm strain_{S_1}}\approx \sum_{n,k}[g_{n,k}j_+a_{n,k}+\rm{H.c.}],
\label{eq:eight}
\end{equation}
where $j_+=j_-^\dagger=\vert 3\rangle_j\langle 1\vert+\vert 4\rangle_j\langle2\vert$ is the spin-conserving raising operator, and $g_{n,k}$ describes the coupling strength with the form
\begin{equation}
g_{n,k}=d\sqrt{\dfrac{\hbar k^2}{2\rho V\omega_{n,k}}}\xi_{n,k}(y,z), 
\end{equation}
where $d/2\pi\approx1$~PHz is the strain sensitivity and the dimensionless function $\xi_{n,k}(y,z)$ denotes specific strain distribution~\cite{Lemonde2018,Meesala2018}.  For a small quasi 1D diamond acoustic waveguide, the length, width, and thickness can be chosen as $L\sim10-100~\mu$m, ~$w,d \lesssim 200$~nm$^2$, respectively, and it has the group velocity $\upsilon\sim1\times10^4~\rm{m/s}$, thus the strain-induced coupling strength is $ g=2\pi\times(4\sim14)$~MHz~\cite{Lemonde2018,Meesala2018,Qiao2020}.

Considering that the acoustic modes are well separated from frequency ($\Delta\omega_n\geq2\pi\times50$ MHz) in the waveguide with small size, we could treat the mechanical mode as a single standing wave with $\omega_n\approx2\pi\times46$~GHz for simplicity. Then, the Hamiltonian Eq.~(\ref{eq:one}) can be written as
\begin{equation}\label{eq:1}
\begin{aligned}
H_{{S_1}}=&\sum_j^{N_1}[\omega_B\vert  2\rangle_j \langle 2\vert +\Delta\vert  3\rangle_j \langle 3\vert +(\omega_B+\Delta)\vert  4\rangle_j \langle 4\vert\\&+\frac{\Omega_1(t)}{2}\vert 1\rangle_j \langle 4\vert e^{i\omega_1t}
+\frac{\Omega_2(t)}{2}\vert  2\rangle_j \langle 3\vert e^{i\omega_2t}+\mathrm{H.c.}]\\&+\omega_{n}a^\dagger a+\sum _{j}^{N_1}[ g_{n}^j(\vert 3\rangle_j \langle 1\vert+\vert 4\rangle_j \langle 2\vert)a+\rm{H.c.}].
\end{aligned}
\end{equation}

Performing a unitary transformation with respect to $U=e^{-iH_0t}$, where $H_0=\sum_j[\left({{\omega_n}-{\omega_2}}\right)\vert 2_j\rangle\langle2\vert+\omega_n\vert 3_j\rangle\langle3\vert+{\omega_1}\vert 4_j\rangle\langle4 \vert+{\omega_{n}}{a^\dag}a]$, the Hamiltonian in the interaction picture reads
\begin{equation}\label{eq:mti}
\begin{split}
H_{I_{S_1}}=&\sum_j^{{N_1}}[(\nu-\delta_2)\vert  2\rangle_j\langle 2\vert+{\nu}\vert 3\rangle_j\langle3\vert +\delta_1\vert  4\rangle_j \langle 4\vert\\
&+\dfrac{\Omega_1(t)}{2}\vert  1\rangle_j \langle 4\vert
+\dfrac{\Omega_2(t)}{2}\vert  2\rangle_j \langle 3\vert\\
&+g_n(\vert  3\rangle_j \langle 1\vert+\vert  4\rangle_j \langle 2\vert)a e^{iw_1t}+\rm{H.c.}],
\end{split}
\end{equation}
where $\nu,\delta_1,\delta_2$ are the corresponding detunings between the frequencies $\omega_{n,k}$, $\omega_1$, $\omega_2$ and eigenfrequencies of states $\vert 3\rangle,\vert 4\rangle$, as shown in Fig.~\ref{fig1}(b), and $w_1=\omega_1 +\omega_2-\omega_n$. Under the condition, $\delta_{1,2}\gg\Omega_{1,2}$,  we may further eliminate the higher energy levels $\vert 3\rangle$ and $\vert 4\rangle$ via Froehlich-Nakajima transformation~\cite{Fr1950,Nakajima1955,Schrieffer1966}. Finally, we obtain an equivalent two-level Hamiltonian
\begin{equation}\label{eq:14}
H_{\rm eq_{S_1}} = \sum_j^{N_1}[\varepsilon_1 \vert 2\rangle_j\langle 2 \vert 
	+\lambda_1 a\vert 2\rangle_j\langle 1\vert  + \Lambda_1 a\vert 1\rangle_j\langle 2\vert e^{iw_1t} +\rm{H.c.} ],
\end{equation}
where the parameters $ \varepsilon_1, \lambda_1, \Lambda_1 $in Eq.~(\ref{eq:14}) have the forms as following
\begin{equation}\label{e1}
\begin{split}
&\varepsilon_1=\nu-\delta_2-\dfrac{\Omega_2^2}{4\delta_2}+\dfrac{\Omega_1^2}{4\delta_1},\\
&\lambda_1=-\left(\dfrac{1}{\nu}+\dfrac{1}{\delta_2}\right)\cdot\dfrac{\Omega_2g_n}{4},\\
&\Lambda_1=-\left(\dfrac{1}{\delta_1}+\dfrac{1}{\delta_1+\delta_2-\nu}\right)\cdot\dfrac{\Omega_1g_n}{4}.
\end{split}
\end{equation}

Next we consider the total hybrid system with two segments $S_{1,2}$, which are connected by a common acoustic mode. The effective Hamiltonian of this whole hybrid system can be written as 
\begin{equation}\label{eq:15}
\begin{split}
H =&\dfrac{\varepsilon_1}{2}J_1^z +\lambda_1aJ_1^+ +\Lambda_1aJ_1^-e^{iw_1t}\\
+&\dfrac{\varepsilon_2}{2}J_2^z +\lambda_2aJ_2^+ +\Lambda_2aJ_2^-e^{iw_2t} + \rm{H.c.} ,
\end{split}
\end{equation}
where the operators $ J_{1,2}^{z}=\sum_{j=1}^{N_{1,2}} (\vert 2\rangle_j\langle 2 \vert -\vert 1\rangle_j\langle 1 \vert), J_{1,2}^{+}=\sum_{j=1}^{N_{1,2}} \vert 2\rangle_j\langle 1 \vert, J_{1,2}^{-}=\sum_{j=1}^{N_{1,2}} \vert 1\rangle_j\langle 2 \vert $  are the collective spin operators of the SiV centers, subscripts 1 and 2 denote the two parts and $ N_1,N_2 $ are the total spin numbers of corresponding SiV-center segments. Moreover, the parameters $ \varepsilon_2, \lambda_2, \Lambda_2 $ in Eq.~(\ref{eq:15}) have the forms as following
\begin{equation}\label{e2}
\begin{split}
&\varepsilon_2=\nu-\delta_4-\dfrac{\Omega_4^2}{4\delta_4}+\dfrac{\Omega_3^2}{4\delta_3},\\
&\lambda_2=-\left(\dfrac{1}{\nu}+\dfrac{1}{\delta_4}\right)\cdot\dfrac{\Omega_4g_n}{4},\\
&\Lambda_2=-\left(\dfrac{1}{\delta_3}+\dfrac{1}{\delta_3+\delta_4-\nu}\right)\cdot\dfrac{\Omega_3g_n}{4}.
\end{split}
\end{equation}

As shown in Eq.~(\ref{e1}), Eq.~(\ref{e2}), by properly adjusting the microwave fields, the effective detunings can be set as $ \varepsilon_1=-\varepsilon_2=\Delta_s$, which implies that the two parts of the SiV centers are physically different. In addition, we set $w_1=w_2=0$. 

Assuming that the setup works at 100mK temperature, thus the phonon number of acoustic mode is close to 0, i.e. $\left\langle a^\dag a\right\rangle\sim0$. Moreover, with the condition $\Delta_s\gg\lambda_{1,2}g_n,\Lambda_{1,2}g_n$ is satisfied, applying the  canonical transformation $H \rightarrow e^{-S}H_{eff}e^S$ with~\cite{Schrieffer1966,Sergey2011} 
\begin{equation}\label{eq:16}
\begin{split}
S=&\dfrac{1}{\Delta_s}\cdot g_n[\lambda_1(a^\dag J_1^- - aJ_1^+)+\Lambda_1(aJ_1^- - a^\dag J_1^+)\\&+\lambda_2(aJ_2^+ - a^\dag J_2^- )+\Lambda_2(a^\dag J_2^+ - aJ_2^-)] .
\end{split}
\end{equation}
Finally, we could obtain an effective projected Hamiltonian in the spin-ensemble subspace~\cite{Sergey2011,Blais2007} as following
\begin{equation}\label{eq:17}
\begin{split}
H_{eff} = \dfrac{\Delta _s}{2}&J_1^z - \dfrac{\Delta _s}{2}J_2^z \\
+\dfrac{1}{\Delta _s} &\cdot \left\lbrace \lambda_1^2 J_1^+J_1^- - \lambda_2^2 J_2^+J_2^- \right. \\
&- \Lambda_1^2J_1^-J_1^+ + \Lambda_2^2J_2^-J_2^+\\
&\left. +(\lambda_1\Lambda_2-\lambda_2\Lambda_1)(J_1^+J_2^+ + J_1^-J_2^-)\right\rbrace .
\end{split}
\end{equation}
\\
Here the terms in lines 2 and 3 represent the OAT interaction, while the term in line 4 indicates the TATS interaction~\cite{Kuzmich1997,Wineland1994,Ma2011}. Thus, by tunning the driving fields, one could realize an OAT Hamiltonian along the z axis, a TATS Hamilatonian, and a mixed Hamiltonian containing the OAT and TATS interactions, respectively. In addition, the dynamical evolution of the system can be described by the quantum master equation
\begin{equation}\label{eq:18}
\begin{split}
\dfrac{d\rho(t)}{dt}=&-i\left[H_{eff},\rho\right]+(n_{th}+1)\varGamma_{eff}D\left[ J_-\right]\rho(t)\\&+n_{th}\varGamma_{eff}D\left[ J_+\right]\rho(t),
\end{split}
\end{equation}
where $n_{th}$ is the average thermal phonon number, and $D\left[\hat{o}\right]\hat{\rho}=\hat{o}\hat{\rho}\hat{o}^\dag-\hat{o}^\dag\hat{o}\hat{\rho}/2-\hat{\rho}\hat{o}^\dag\hat{o}/2$ is the standard Linblad superoperator. $\varGamma_{eff}=\varGamma_m\cdot(\lambda_{1,2}/\Delta_s)^2$ indicates the collective spin relaxation induced by mechanical dissipation $\varGamma_m$ of the corresponding acoustic mode.

\section{Spin squeezing}
In this section, we quantify the degree of spin-squeezed states by calculating two most frequently used squeezing parameters, $\xi_S^2$ and $\xi_R^2$. $\xi_S^2$ is closely linked to quantum correlations (entanglement), whereas $\xi_R^2$ finds wide applications in quantum metrology. These two parameters are defined as follows~\cite{Kuzmich1997,Wineland1994,Wineland1992,Ma2011}
\begin{equation}\label{eq:19}
\begin{split}
&\xi_S^2=\dfrac{4(\Delta J_{\overrightarrow{n}_\bot})^2_{min}}{N_{tot}}\\
&\xi_R^2=\dfrac{N_{tot}(\Delta J_{\overrightarrow{n}_\bot})^2_{min}}{\left|\overrightarrow{\langle J\rangle}\right|^2 },
\end{split}
\end{equation}
where $(\Delta J_{\overrightarrow{n}_\bot})^2_{min}$ is the minimum variance in a direction which perpendicular to the mean spin direction, and $ \left|\overrightarrow{J}\right|=\sqrt{\left\langle J_x^2\right\rangle +\left\langle J_y^2\right\rangle+\left\langle J_z^2\right\rangle}$ denotes the magnitude of the mean spin. $ N_{tot}=N_1+N_2 $ is the total number of SiV centers in the waveguide, and for the sake of simplicity, we assume that $N_1\simeq N_2$.

\subsection{OAT interaction Hamiltonian}
When the terms in Eq.~(\ref{eq:17}) are set as $\lambda_1\Lambda_2-\lambda_2\Lambda_1=0$ and $\lambda_1^2\neq\Lambda_1^2,\lambda_2^2\neq\Lambda_2^2$ through tunning the amplitudes and frequencies of the driving fields, we can obtain the following OAT Hamiltonian along the z axis,
\begin{equation}\label{eq:20}
\begin{split}
H_{OAT}=\dfrac{\Delta_{s1}}{2}J_1^z-\dfrac{\Delta_{s2}}{2}J_2^z + G_{OAT1}J_1^+J_1^-+G_{OAT2}J_2^+J_2^-,
\end{split}
\end{equation}
where $G_{OAT~1,2}=(\lambda_{1,2}^2-\Lambda_{1,2}^2)/\Delta_s$ describe the OAT interaction strengths of corresponding spin ensembles, and $\Delta_{s1,s2}=\Delta_{s}+2\rm min \left\lbrace \lambda_{1,2}^2,\Lambda_{1,2}^2\right\rbrace/\Delta_s $. Considering that $J_+J_-=J^2-J_z^2+J_z$, the last two terms in Eq.~(\ref{eq:17}) indicate a standard OAT interaction Hamiltonian~\cite{Ma2011}. 
%=================================================================== 
\begin{figure}[tpb]
\centering
\includegraphics[width=1\linewidth]{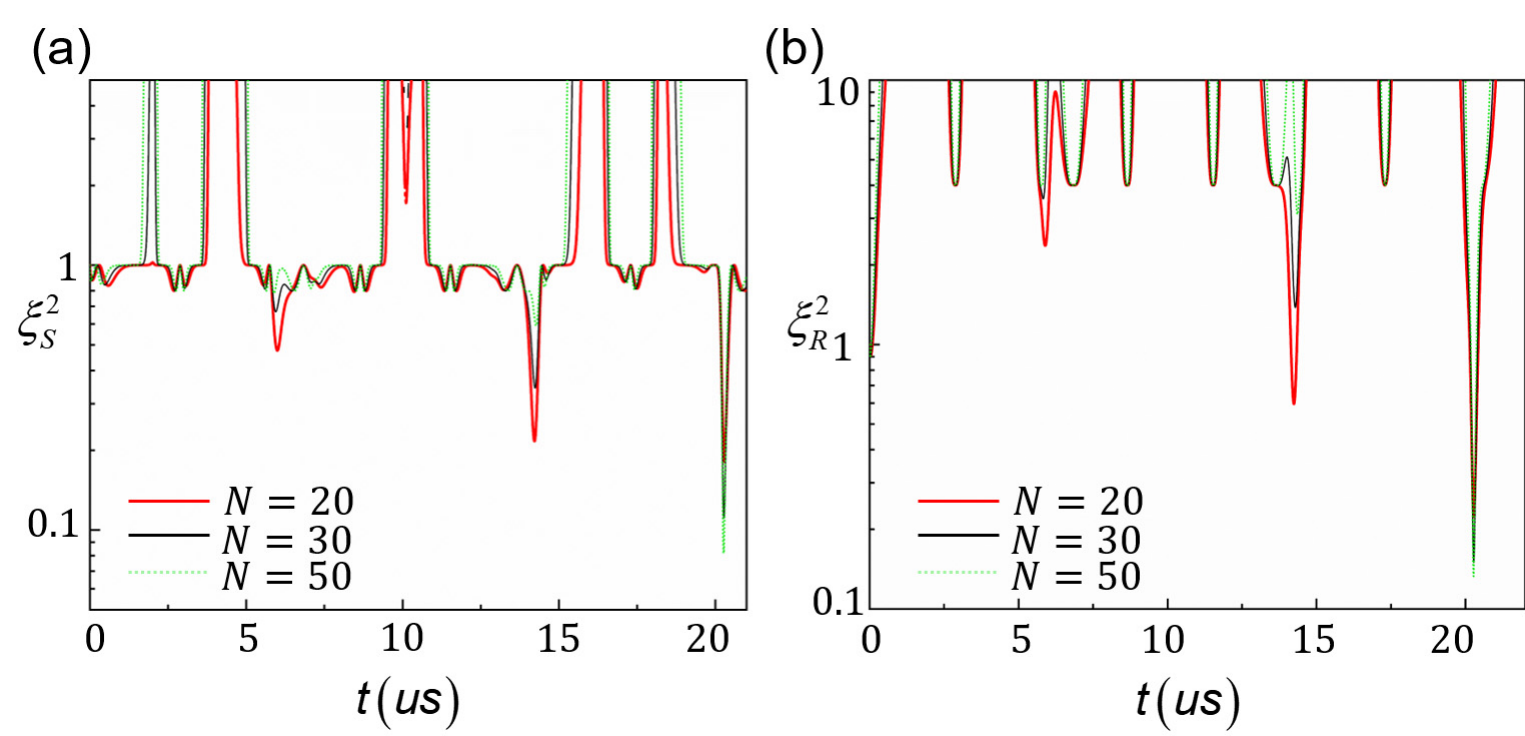}
\caption{The effects of spin number to the spin squeezing generated by the OAT interaction. (a) Time evolution of the squeezing parameter $\xi_S^2$ with $N_1=N_2=N=20,30,50$. (b) Time evolution of the squeezing parameter $\xi_R^2$ with $N_1=N_2=N=20,30,50$.}
\label{fig2}
\end{figure}
%=================================================================== 

Figure~\ref{fig2} shows the time evolution of the squeezing parameters $\xi_S^2$ and $\xi_R^2$. The red line, black line, and green dotted line represent the squeezing parameters of $N_1=N_2=N=20,30,50$, respectively. The hybrid system evolves from a spin coherent state distributed on the x-axis, in which state the values of both parameters $\xi_S^2$ and $\xi_R^2$ are 1, as depicted in the figure~\ref{fig2} at time 0. As the system begins to evolve, these two parameters become smaller than 1, indicating that the spin squeezing has been generated in this hybrid system. As shown in figure~\ref{fig2}, the generated spin-squeezed states reaches their optimal squeezing at time $t\sim20us$, with minimum values are $\xi_S^2\approx0.18,0.11,0.08$ and $\xi_R^2\approx0.22,0.15,0.13$ with the case of $N=20,30,50$, respectively. In addition, we find that $\xi_S^2<\xi_R^2$ for the same spin numbers, which is consistent with the results as mentioned in Ref.~\cite{Ma2011}.

\subsection{TATS interaction Hamiltonian}
Similar to the case of OAT interaction, we can also set $\lambda_1\Lambda_2-\lambda_2\Lambda_1\neq0$ and $\lambda_1^2=\Lambda_1^2,\lambda_2^2=\Lambda_2^2$ by tunning the amplitudes and frequencies of the driving fields. Then, the Hamiltonian with a TATS interaction could be obtained from Eq.~(\ref{eq:17}) as follows
\begin{equation}\label{eq:21}
\begin{split}
H_{TATS}&=\dfrac{\Delta_{s1}}{2}J_1^z-\dfrac{\Delta_{s2}}{2}J_2^z + G_{TATS}(J_1^+J_2^+ + J_1^-J_2^-),
\end{split}
\end{equation}
where $G_{TATS}=(\lambda_1\Lambda_2-\lambda_2\Lambda_1)/\Delta_s$ indicates the TATS interaction strength, and $\Delta_{s1,s2}=\Delta_{s}+2\lambda_{1,2}^2/\Delta_s $.
%=================================================================== 
\begin{figure}[tpb]
\centering
\includegraphics[width=1\linewidth]{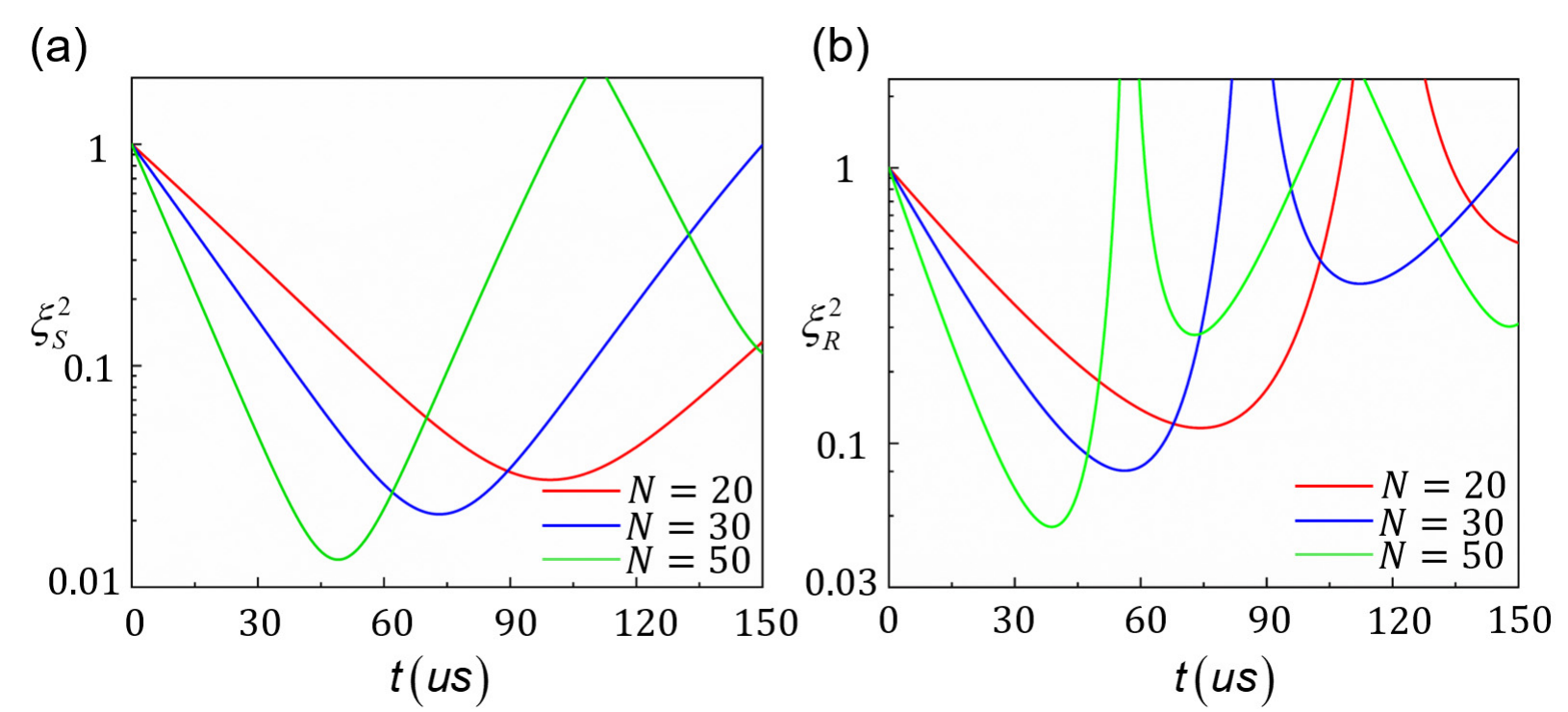}
\caption{The effects of spin number to the spin squeezing generated by the TATS interaction. (a) Time evolution of the squeezing parameter $\xi_S^2$ with $N_1=N_2=N=20,30,50$. (b) Time evolution of the squeezing parameter $\xi_R^2$ with $N_1=N_2=N=20,30,50$.}
\label{fig3}
\end{figure}
%=================================================================== 

Figure~\ref{fig3} depicts the time evolution of squeezing parameters $\xi_S^2$ and $\xi_R^2$ with different spin numbers in the case of TATS interaction Hamiltonian. The red, blue, green lines in this figure represent the squeezing parameters $\xi_S^2$ and $\xi_R^2$ of $N_1=N_2=N=20,30,50$, respectively. We can see that the minimum values of $\xi_S^2$ and $\xi_R^2$ have decreased significantly compared to the OAT interaction case in fig.~\ref{fig3} with the same spin number, specifically, $\xi_S^2\approx0.03,0.021,0.013$ and $\xi_R^2\approx0.113,0.079,0.049$ with the case of $N=20,30,50$, respectively. Similarly, $\xi_S^2<\xi_R^2$ for the same spin numbers. Figure~\ref{fig3} shows that both squeezing parameters would reach their minimum values more quickly with the increment of the spin numbers. Moreover, we can see that the spin-squeezed state generated by this TATS interaction Hamiltonian could approach the Heisenberg limit $1/N$ for large spin numbers, which is not possible in the OAT case~\cite{Ma2011}.  

\subsection{Mixed Hamiltonian of OAT and TATS interaction }

With appropriate tuning of the microwave driving fields, it is also possible to obtain a mixed Hamiltonian that comprises both OAT and TATS interactions from Eq.~(\ref{eq:17}),
\begin{equation}\label{eq:22}
\begin{split}
H_{mix}=&\dfrac{\Delta_{s1}}{2}J_1^z-\dfrac{\Delta_{s2}}{2}J_2^z \\&+ G_{mix}(J_1^-J_1^+ + J_2^+J_2^- + J_1^+J_2^+ + J_1^-J_2^-),
\end{split}
\end{equation}
where $G_{mix}$ represents the mixed interaction strength, and $\Delta_{s1,s2}=\Delta_{s}+2 \rm min \left\lbrace \lambda_{1,2}^2,\Lambda_{1,2}^2\right\rbrace/\Delta_s $.
%=================================================================== 
\begin{figure}
\centering
\includegraphics[width=1\linewidth]{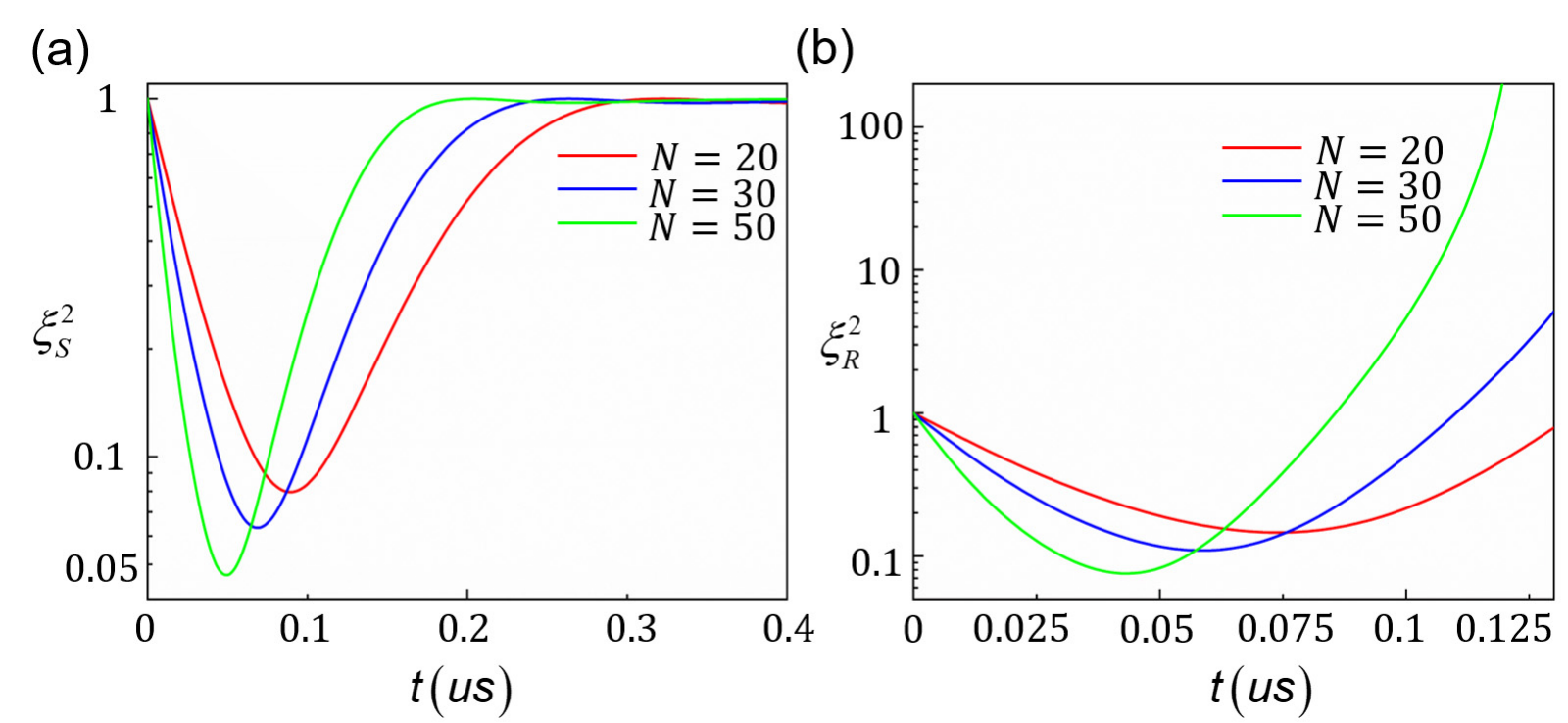}
\caption{The effects of spin number to the spin squeezing generated by the mixed Hamiltonian. (a) Time evolution of the squeezing parameter $\xi_S^2$ with $N_1=N_2=N=20,30,50$. (b) Time evolution of the squeezing parameter $\xi_R^2$ with $N_1=N_2=N=20,30,50$.}
\label{fig4}
\end{figure}
%=================================================================== 

We also plotted the time evolution of the squeezing parameters $\xi_S^2$ and $\xi_R^2$ for different spin numbers in Fig.~\ref{fig4}, and the black, red, green lines in this figure represent the squeezing parameters $\xi_S^2$ and $\xi_R^2$ of $N_1=N_2=N=20,30,50$, respectively. In the case of mixed Hamiltonian of the OAT and TATS interaction, the minimum values of corresponding squeezing parameters are $\xi_S^2\approx0.08,0.063,0.047$ and $\xi_R^2\approx0.146,0.109,0.075$ with the case of $N=20,30,50$, respectively, which are smaller than the case of OAT interaction induced spin squeezing, but also slightly larger than the ideal TATS case. From the Fig.~\ref{fig4}, we can also see that the time for the system to reach the optimal squeezing is significantly smaller than the OAT (Fig.~\ref{fig2}) and TATS (Fig.~\ref{fig3}) cases. 

\begin{figure}
\centering
\includegraphics[width=1\linewidth]{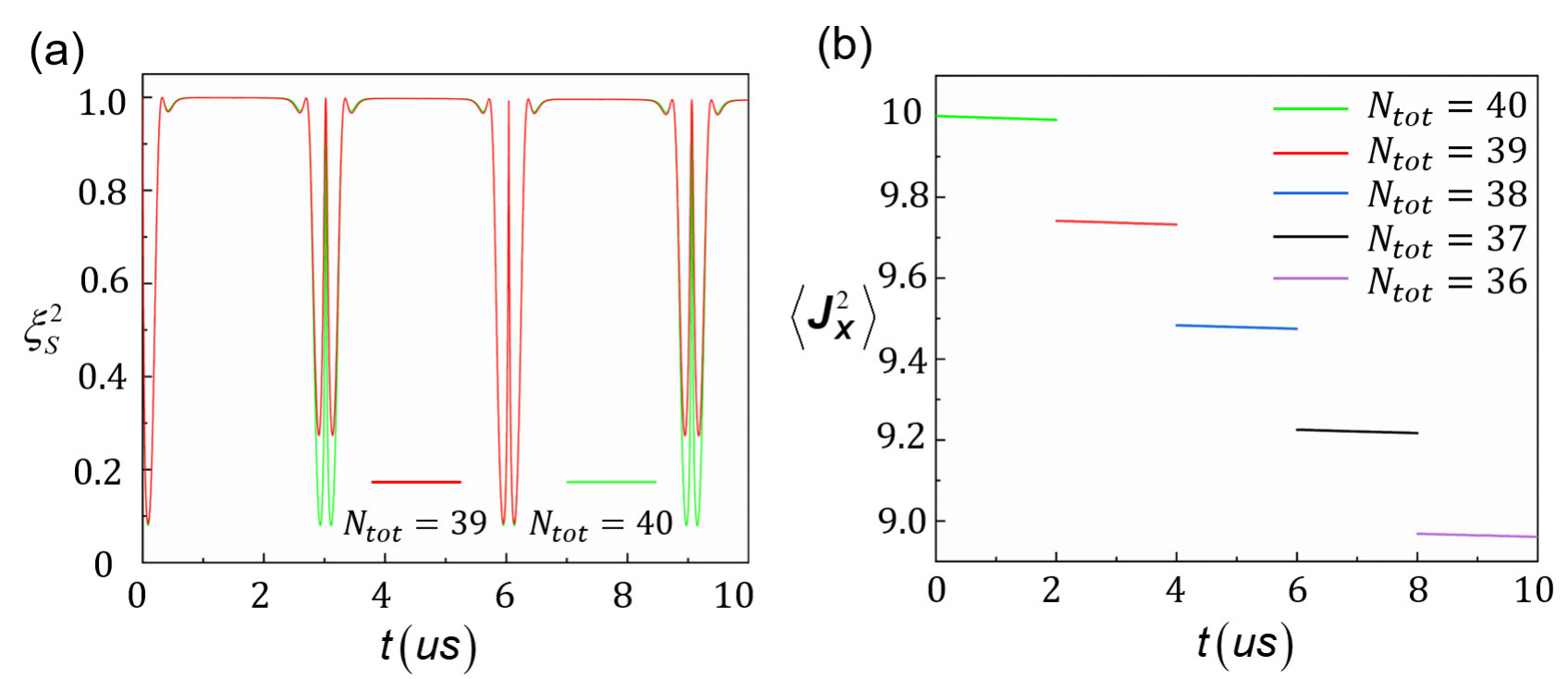}
\caption{The effects of even and odd total spin numbers to the spin squeezing generated by the mixed Hamiltonian. (a) Time evolution of the squeezing parameter $\xi_S^2$ with $N_{tot}=40$ and $N_{tot}=39$. (b) An example of how the even-odd sensitivity of the spin squeezing could be used for sensing. Time evolution of the value of $\left\langle J_X^2 \right\rangle$ with $N_{tot}=40,39,38,37,36$.}
\label{fig5}
\end{figure}

In particular, in the mixed OAT-TATS interaction case, we find that the spin squeezing effect differs significantly depending on whether the total number of spins is odd or even. This property may be utilized to detect changes of the number, $N_{tot}$, of coupled spins at the single-particle level. Figure~\ref{fig5} shows the time evolution of squeezing parameters $\xi_S^2$ with $N_{tot}=40$ and $N_{tot}=39$. Notably, during the first instance of spin squeezing, the parameters $\xi_S^2$ of $N_{tot}=40$ and $N_{tot}=39$ are almost identical. However, as the hybrid evolves from the spin coherent state to spin-squeezed state for the second time, the spin squeezing in the $N_{tot}=39$ case is significantly poorer compared to the the $N_{tot}=40$ case, as shown in Fig.~\ref{fig5}(a). When the number of total spins is odd, there will be a difference in the parity of spin numbers between the two segments, resulting in the overall dynamics of spin squeezing, the combination of two parts with different periods and parities. Therefore, like destructive and constructive interference, the squeezing parameters $\xi_S^2$ with odd total spins will display the maximum squeezed value that alternates between large and small in odd and even periods. In contrast, in the case with even total spins, such alternations in the maximum value of spin squeezing are absent. Figure~\ref{fig5}(b) illustrates that how this odd-even sensitivity could be used for sensing. We plot the value of $\left\langle J_X^2 \right\rangle$ with different total spin numbers $N_{tot}=40,39,38,37,36$. When the spins leaving or decoupling from the waveguide one by one, the corresponding values of $\left\langle J_X^2 \right\rangle$ would become smaller and smaller, as shown in Fig.~\ref{fig5}(b). 

%=================================================================== 
%=================================================================== 

\section{Experimental feasibility}

%=================================================================== 
\begin{figure}[tpb]
\centering
\includegraphics[width=0.9\linewidth]{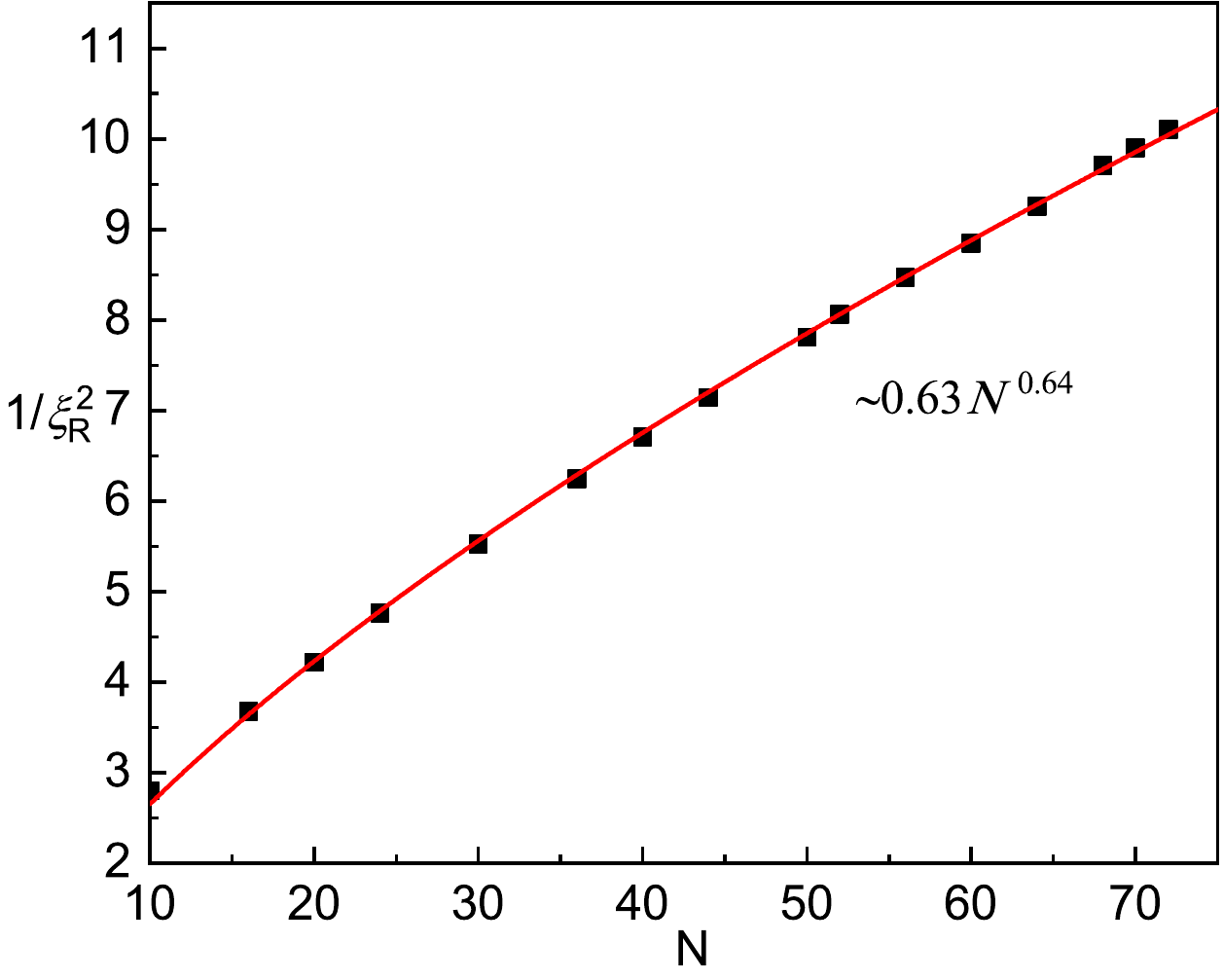}
\caption{The optimal squeezing parameter $(1/\xi_R^2)_{max}$ versus the total spin numbers $N_{tot}$ with the consideration of experimental disspation in TATS interaction case. The black dots is the value of $(1/\xi_R^2)_{max}$ with corresponding spin numbers, and the red line represents the curve fitting of numerical results. }
\label{fig6}
\end{figure}
%=================================================================== 

In this section, we discuss the relevant parameters used in numerical simulations to assess the practical feasibility of this schemement. First, we take the value of the strain-induced coupling strength between the acoustic mode in the diamond waveguide and SiV centers  to $g=2\pi\times5\rm{MHz}$. To embed the SiV centers into the 1D diamond waveguide, we can utilize ion implantation techniques based on state-of-the-art nanofabrication techniques~\cite{WangJunfeng2017}. The ground state splitting of SiV centers is $\Delta \approx $ 46GHz, and the transitions between states $\left| 1\right\rangle  \leftrightarrow\left| 4\right\rangle$ and $\left| 2\right\rangle  \leftrightarrow\left| 3\right\rangle$ can be induced by the microwave driving fields or via an equivalent optical Raman process, which has already been experimentally realized~\cite{Lemonde2018,Jarryd2012,Pingault2014}.  At 100 mK, the spin dephasing rate of a single  SiV center is about $ \gamma_{d}\sim 100 $Hz, corresponding to a coherent time is $ T_{s}\sim $10ms~\cite{Sipahigil2014,Sipahigil2016,Sukachev2017}. The driving fields adopted here are $\Omega_1,\Omega_2,\Omega_3,\Omega_4\sim 2\pi\times(30\sim50)\rm{MHz}$ with $\delta_1,\delta_2,\delta_3,\delta_4\sim 2\pi\times(300\sim500)\rm{MHz}$, respectively. It should be noted that we have not taken into account the effect of dissipation in numerical simulations of the squeezing in the previous section. A quality factor of $Q\approx\times10^5$ for the mechanical phonon modes of the small-sized diamond waveguide has been demonstrated~\cite{Khanaliloo2015,Ovartchaiyapong2012}, which leads to a mechanical dissipation value of $\varGamma\sim2\pi\times500~\rm kHz$. Consequently, the effective collective decay rate induced by mechanical dissipation in Eq.~(\ref{eq:18}) has a value as $\varGamma_{eff}\sim2\pi\times50\rm{Hz}$ in the TATS interaction case, which is of the same order of magnitude as the spin dephasing rate. Here, we modify the master equation Eq.~(\ref{eq:18}) by including the spin dephasing term as follows
\begin{equation}\label{eq:23}
\begin{split}
\dfrac{d\rho(t)}{dt}=&-i\left[H_{eff},\rho\right]+(n_{th}+1)\varGamma_{eff}D\left[ J_-\right]\rho(t)\\&+n_{th}\varGamma_{eff}D\left[ J_+\right]\rho(t)+ \gamma_d\textstyle\sum_jD\left[ \sigma_z^j\right]\rho(t).
\end{split}
\end{equation}

In Fig.\ref{fig6}, we plot the optimal squeezing parameter $(1/\xi_R^2)_{max}$ versus the total number of spins $N_{tot}$ in TATS interaction case, taking into account the collective decoherence induced by mechanical dissipation of the acoustic mode and the dephasing rate of SiV centers. Using the numriacl results obtained from Eq.~(\ref{eq:23}), we fit a curve and obtain the trend of the optimal squeezing parameters $(\xi_R^2)_{max}$ with respect to the number of spins as $\xi_R^2\sim1.61N^{-0.64}$. As such, our scheme can generate highly squeezed-spin states under currently available experimental conditions in the hybrid system base on SiV centers.

\section{Conclusion}

In summary, we have designed a hybrid quantum system, consisting of an ensemble of SiV centers coupled to the acoustic mode of a diamond waveguide via the strain-induced coupling. The system is partitioned into two segments with different sets of microwave driving fields, and by ajusting the frequencies and amplitudes of fields, we can achieve the OAT interaction, the TATS interaction and mixed Hamiltonian with both OAT and TATS interactions. The scheme can still work when the numbers of SiV centers in the two segments differ, despite a reduction in the squeezing effect. In the ideal TATS scenario with large numbers of spins, the two spin-squeezing parameters $\xi_R^2$ and $\xi_S^2$ scale with total spin numbers as $\xi_S^2,\xi_R^2\sim N^{-1}$, reaching the Heisenberg limit. In the mixed interaction case, our hybrid system can generate the optimal spin squeezing more rapidly, and these spin-squeezed states is sensitive to the parity of the total number of spins. Moreover, we have provided a possible method for measuring the change of spin numbers at the single particle level. Considering the realistic dissipation, $\xi_R^2$ scales with the total number of spins as $\xi_R^2\sim1.61N^{-0.64}$, demonstrating its potential for application in quantum metrology. Consequently, our scheme can work well under experimental conditions and extend the applications of the SiV-based hybrid quantum systems in quantum information processing and quantum metrology.  

%=================================================================== 
%=================================================================== 

\section{Acknowledgments}

This work was supported by the National Key R\&D Program of China (Grant No. 2019YFA0308200) and the National Natural Science Foundation of China (Grant No. 11874432). We thank Yuan Zhou for valuable discussions.

%=================================================================== 
%=================================================================== 

\bibliography{reference}

%apsrev4-2.bst 2019-01-14 (MD) hand-edited version of apsrev4-1.bst
%Control: key (0)
%Control: author (8) initials jnrlst
%Control: editor formatted (1) identically to author
%Control: production of article title (0) allowed
%Control: page (0) single
%Control: year (1) truncated
%Control: production of eprint (0) enabled
\begin{thebibliography}{80}%
\makeatletter
\providecommand \@ifxundefined [1]{%
 \@ifx{#1\undefined}
}%
\providecommand \@ifnum [1]{%
 \ifnum #1\expandafter \@firstoftwo
 \else \expandafter \@secondoftwo
 \fi
}%
\providecommand \@ifx [1]{%
 \ifx #1\expandafter \@firstoftwo
 \else \expandafter \@secondoftwo
 \fi
}%
\providecommand \natexlab [1]{#1}%
\providecommand \enquote  [1]{``#1''}%
\providecommand \bibnamefont  [1]{#1}%
\providecommand \bibfnamefont [1]{#1}%
\providecommand \citenamefont [1]{#1}%
\providecommand \href@noop [0]{\@secondoftwo}%
\providecommand \href [0]{\begingroup \@sanitize@url \@href}%
\providecommand \@href[1]{\@@startlink{#1}\@@href}%
\providecommand \@@href[1]{\endgroup#1\@@endlink}%
\providecommand \@sanitize@url [0]{\catcode `\\12\catcode `\$12\catcode
  `\&12\catcode `\#12\catcode `\^12\catcode `\_12\catcode `\%12\relax}%
\providecommand \@@startlink[1]{}%
\providecommand \@@endlink[0]{}%
\providecommand \url  [0]{\begingroup\@sanitize@url \@url }%
\providecommand \@url [1]{\endgroup\@href {#1}{\urlprefix }}%
\providecommand \urlprefix  [0]{URL }%
\providecommand \Eprint [0]{\href }%
\providecommand \doibase [0]{https://doi.org/}%
\providecommand \selectlanguage [0]{\@gobble}%
\providecommand \bibinfo  [0]{\@secondoftwo}%
\providecommand \bibfield  [0]{\@secondoftwo}%
\providecommand \translation [1]{[#1]}%
\providecommand \BibitemOpen [0]{}%
\providecommand \bibitemStop [0]{}%
\providecommand \bibitemNoStop [0]{.\EOS\space}%
\providecommand \EOS [0]{\spacefactor3000\relax}%
\providecommand \BibitemShut  [1]{\csname bibitem#1\endcsname}%
\let\auto@bib@innerbib\@empty
%</preamble>
\bibitem [{\citenamefont {Kitagawa}\ and\ \citenamefont
  {Ueda}(1993)}]{Kitagawa1993}%
  \BibitemOpen
  \bibfield  {author} {\bibinfo {author} {\bibfnamefont {M.}~\bibnamefont
  {Kitagawa}}\ and\ \bibinfo {author} {\bibfnamefont {M.}~\bibnamefont
  {Ueda}},\ }\bibfield  {title} {\bibinfo {title} {Squeezed spin states},\
  }\href {https://doi.org/10.1103/PhysRevA.47.5138} {\bibfield  {journal}
  {\bibinfo  {journal} {Phys. Rev. A}\ }\textbf {\bibinfo {volume} {47}},\
  \bibinfo {pages} {5138} (\bibinfo {year} {1993})}\BibitemShut {NoStop}%
\bibitem [{\citenamefont {Wineland}\ \emph {et~al.}(1994)\citenamefont
  {Wineland}, \citenamefont {Bollinger}, \citenamefont {Itano},\ and\
  \citenamefont {Heinzen}}]{Wineland1994}%
  \BibitemOpen
  \bibfield  {author} {\bibinfo {author} {\bibfnamefont {D.~J.}\ \bibnamefont
  {Wineland}}, \bibinfo {author} {\bibfnamefont {J.~J.}\ \bibnamefont
  {Bollinger}}, \bibinfo {author} {\bibfnamefont {W.~M.}\ \bibnamefont
  {Itano}},\ and\ \bibinfo {author} {\bibfnamefont {D.~J.}\ \bibnamefont
  {Heinzen}},\ }\bibfield  {title} {\bibinfo {title} {Squeezed atomic states
  and projection noise in spectroscopy},\ }\href
  {https://doi.org/10.1103/PhysRevA.50.67} {\bibfield  {journal} {\bibinfo
  {journal} {Phys. Rev. A}\ }\textbf {\bibinfo {volume} {50}},\ \bibinfo
  {pages} {67} (\bibinfo {year} {1994})}\BibitemShut {NoStop}%
\bibitem [{\citenamefont {Ma}\ \emph {et~al.}(2011)\citenamefont {Ma},
  \citenamefont {Wang}, \citenamefont {Sun},\ and\ \citenamefont
  {Nori}}]{Ma2011}%
  \BibitemOpen
  \bibfield  {author} {\bibinfo {author} {\bibfnamefont {J.}~\bibnamefont
  {Ma}}, \bibinfo {author} {\bibfnamefont {X.}~\bibnamefont {Wang}}, \bibinfo
  {author} {\bibfnamefont {C.}~\bibnamefont {Sun}},\ and\ \bibinfo {author}
  {\bibfnamefont {F.}~\bibnamefont {Nori}},\ }\bibfield  {title} {\bibinfo
  {title} {Quantum spin squeezing},\ }\href
  {https://doi.org/https://doi.org/10.1016/j.physrep.2011.08.003} {\bibfield
  {journal} {\bibinfo  {journal} {Phys. Rep.}\ }\textbf {\bibinfo {volume}
  {509}},\ \bibinfo {pages} {89} (\bibinfo {year} {2011})}\BibitemShut
  {NoStop}%
\bibitem [{\citenamefont {Sørensen}\ \emph {et~al.}(2001)\citenamefont
  {Sørensen}, \citenamefont {Duan}, \citenamefont {Cirac},\ and\ \citenamefont
  {Zoller}}]{Zoller2001}%
  \BibitemOpen
  \bibfield  {author} {\bibinfo {author} {\bibfnamefont {A.}~\bibnamefont
  {Sørensen}}, \bibinfo {author} {\bibfnamefont {L.}~\bibnamefont {Duan}},
  \bibinfo {author} {\bibfnamefont {J.}~\bibnamefont {Cirac}},\ and\ \bibinfo
  {author} {\bibfnamefont {P.}~\bibnamefont {Zoller}},\ }\bibfield  {title}
  {\bibinfo {title} {Many-particle entanglement with bose–einstein
  condensates},\ }\href {https://www.nature.com/articles/35051038/} {\bibfield
  {journal} {\bibinfo  {journal} {Nature}\ }\textbf {\bibinfo {volume} {409}},\
  \bibinfo {pages} {63} (\bibinfo {year} {2001})}\BibitemShut {NoStop}%
\bibitem [{\citenamefont {Sørensen}\ and\ \citenamefont
  {Mølmer}(2001)}]{Anders2001}%
  \BibitemOpen
  \bibfield  {author} {\bibinfo {author} {\bibfnamefont {A.~S.}\ \bibnamefont
  {Sørensen}}\ and\ \bibinfo {author} {\bibfnamefont {K.}~\bibnamefont
  {Mølmer}},\ }\bibfield  {title} {\bibinfo {title} {Entanglement and extreme
  spin squeezing},\ }\href {https://doi.org/10.1103/PhysRevLett.86.4431}
  {\bibfield  {journal} {\bibinfo  {journal} {Phys. Rev. Lett.}\ }\textbf
  {\bibinfo {volume} {86}},\ \bibinfo {pages} {4431} (\bibinfo {year}
  {2001})}\BibitemShut {NoStop}%
\bibitem [{\citenamefont {Cirac}\ and\ \citenamefont
  {Zoller}(1995)}]{Cirac1995}%
  \BibitemOpen
  \bibfield  {author} {\bibinfo {author} {\bibfnamefont {J.~I.}\ \bibnamefont
  {Cirac}}\ and\ \bibinfo {author} {\bibfnamefont {P.}~\bibnamefont {Zoller}},\
  }\bibfield  {title} {\bibinfo {title} {Quantum computations with cold trapped
  ions},\ }\href {https://doi.org/10.1103/PhysRevLett.74.4091} {\bibfield
  {journal} {\bibinfo  {journal} {Phys. Rev. Lett.}\ }\textbf {\bibinfo
  {volume} {74}},\ \bibinfo {pages} {4091} (\bibinfo {year}
  {1995})}\BibitemShut {NoStop}%
\bibitem [{\citenamefont {Duan}\ \emph {et~al.}(2001)\citenamefont {Duan},
  \citenamefont {Lukin}, \citenamefont {Cirac},\ and\ \citenamefont
  {Zoller}}]{Duan2001}%
  \BibitemOpen
  \bibfield  {author} {\bibinfo {author} {\bibfnamefont {L.}~\bibnamefont
  {Duan}}, \bibinfo {author} {\bibfnamefont {M.}~\bibnamefont {Lukin}},
  \bibinfo {author} {\bibfnamefont {J.}~\bibnamefont {Cirac}},\ and\ \bibinfo
  {author} {\bibfnamefont {P.}~\bibnamefont {Zoller}},\ }\bibfield  {title}
  {\bibinfo {title} {Long-distance quantum communication with atomic ensembles
  and linear optics},\ }\href {https://www.nature.com/articles/35106500}
  {\bibfield  {journal} {\bibinfo  {journal} {Nature}\ }\textbf {\bibinfo
  {volume} {414}},\ \bibinfo {pages} {413} (\bibinfo {year}
  {2001})}\BibitemShut {NoStop}%
\bibitem [{\citenamefont {Kimble}(2008)}]{Kimble2008}%
  \BibitemOpen
  \bibfield  {author} {\bibinfo {author} {\bibfnamefont {H.~J.}\ \bibnamefont
  {Kimble}},\ }\bibfield  {title} {\bibinfo {title} {The quantum internet},\
  }\href {https://www.nature.com/articles/nature07127} {\bibfield  {journal}
  {\bibinfo  {journal} {Nature}\ }\textbf {\bibinfo {volume} {453}},\ \bibinfo
  {pages} {1023} (\bibinfo {year} {2008})}\BibitemShut {NoStop}%
\bibitem [{\citenamefont {Ma}\ \emph {et~al.}(2018)\citenamefont {Ma},
  \citenamefont {Zeng},\ and\ \citenamefont {Zhou}}]{Ma2018}%
  \BibitemOpen
  \bibfield  {author} {\bibinfo {author} {\bibfnamefont {X.}~\bibnamefont
  {Ma}}, \bibinfo {author} {\bibfnamefont {P.}~\bibnamefont {Zeng}},\ and\
  \bibinfo {author} {\bibfnamefont {H.}~\bibnamefont {Zhou}},\ }\bibfield
  {title} {\bibinfo {title} {Phase-matching quantum key distribution},\ }\href
  {https://doi.org/10.1103/PhysRevX.8.031043} {\bibfield  {journal} {\bibinfo
  {journal} {Phys. Rev. X}\ }\textbf {\bibinfo {volume} {8}},\ \bibinfo {pages}
  {031043} (\bibinfo {year} {2018})}\BibitemShut {NoStop}%
\bibitem [{\citenamefont {Khatri}(2021)}]{Khatri2021}%
  \BibitemOpen
  \bibfield  {author} {\bibinfo {author} {\bibfnamefont {S.}~\bibnamefont
  {Khatri}},\ }\bibfield  {title} {\bibinfo {title} {Policies for elementary
  links in a quantum network},\ }\href
  {https://quantum-journal.org/papers/q-2021-09-07-537/} {\bibfield  {journal}
  {\bibinfo  {journal} {Quantum}\ }\textbf {\bibinfo {volume} {5}} (\bibinfo
  {year} {2021})}\BibitemShut {NoStop}%
\bibitem [{\citenamefont {Giovannetti}\ \emph {et~al.}(2006)\citenamefont
  {Giovannetti}, \citenamefont {Lloyd},\ and\ \citenamefont
  {Maccone}}]{Giovannetti2006}%
  \BibitemOpen
  \bibfield  {author} {\bibinfo {author} {\bibfnamefont {V.}~\bibnamefont
  {Giovannetti}}, \bibinfo {author} {\bibfnamefont {S.}~\bibnamefont {Lloyd}},\
  and\ \bibinfo {author} {\bibfnamefont {L.}~\bibnamefont {Maccone}},\
  }\bibfield  {title} {\bibinfo {title} {Quantum metrology},\ }\href
  {https://doi.org/10.1103/PhysRevLett.96.010401} {\bibfield  {journal}
  {\bibinfo  {journal} {Phys. Rev. Lett.}\ }\textbf {\bibinfo {volume} {96}},\
  \bibinfo {pages} {010401} (\bibinfo {year} {2006})}\BibitemShut {NoStop}%
\bibitem [{\citenamefont {Riedel}\ \emph {et~al.}(2010)\citenamefont {Riedel},
  \citenamefont {Böhi}, \citenamefont {Li}, \citenamefont {Hänsch},
  \citenamefont {Sinatra},\ and\ \citenamefont {Treutlein}}]{Max2010}%
  \BibitemOpen
  \bibfield  {author} {\bibinfo {author} {\bibfnamefont {M.~F.}\ \bibnamefont
  {Riedel}}, \bibinfo {author} {\bibfnamefont {P.}~\bibnamefont {Böhi}},
  \bibinfo {author} {\bibfnamefont {Y.}~\bibnamefont {Li}}, \bibinfo {author}
  {\bibfnamefont {T.~W.}\ \bibnamefont {Hänsch}}, \bibinfo {author}
  {\bibfnamefont {A.}~\bibnamefont {Sinatra}},\ and\ \bibinfo {author}
  {\bibfnamefont {P.}~\bibnamefont {Treutlein}},\ }\bibfield  {title} {\bibinfo
  {title} {Atom-chip-based generation of entanglement for quantum metrology},\
  }\href {https://www.nature.com/articles/nature08988} {\bibfield  {journal}
  {\bibinfo  {journal} {Nature}\ }\textbf {\bibinfo {volume} {464}},\ \bibinfo
  {pages} {1170} (\bibinfo {year} {2010})}\BibitemShut {NoStop}%
\bibitem [{\citenamefont {Wang}\ \emph {et~al.}(2018)\citenamefont {Wang},
  \citenamefont {Wang}, \citenamefont {Zhan}, \citenamefont {Bian},
  \citenamefont {Li}, \citenamefont {Sanders},\ and\ \citenamefont
  {Xue}}]{Wang2018}%
  \BibitemOpen
  \bibfield  {author} {\bibinfo {author} {\bibfnamefont {K.}~\bibnamefont
  {Wang}}, \bibinfo {author} {\bibfnamefont {X.}~\bibnamefont {Wang}}, \bibinfo
  {author} {\bibfnamefont {X.}~\bibnamefont {Zhan}}, \bibinfo {author}
  {\bibfnamefont {Z.}~\bibnamefont {Bian}}, \bibinfo {author} {\bibfnamefont
  {J.}~\bibnamefont {Li}}, \bibinfo {author} {\bibfnamefont {B.~C.}\
  \bibnamefont {Sanders}},\ and\ \bibinfo {author} {\bibfnamefont
  {P.}~\bibnamefont {Xue}},\ }\bibfield  {title} {\bibinfo {title}
  {Entanglement-enhanced quantum metrology in a noisy environment},\ }\href
  {https://doi.org/10.1103/PhysRevA.97.042112} {\bibfield  {journal} {\bibinfo
  {journal} {Phys. Rev. A}\ }\textbf {\bibinfo {volume} {97}},\ \bibinfo
  {pages} {042112} (\bibinfo {year} {2018})}\BibitemShut {NoStop}%
\bibitem [{\citenamefont {Wolfgramm}\ \emph {et~al.}(2010)\citenamefont
  {Wolfgramm}, \citenamefont {Cer\`e}, \citenamefont {Beduini}, \citenamefont
  {Predojevi\ifmmode~\acute{c}\else \'{c}\fi{}}, \citenamefont {Koschorreck},\
  and\ \citenamefont {Mitchell}}]{Wolfgramm2010}%
  \BibitemOpen
  \bibfield  {author} {\bibinfo {author} {\bibfnamefont {F.}~\bibnamefont
  {Wolfgramm}}, \bibinfo {author} {\bibfnamefont {A.}~\bibnamefont {Cer\`e}},
  \bibinfo {author} {\bibfnamefont {F.~A.}\ \bibnamefont {Beduini}}, \bibinfo
  {author} {\bibfnamefont {A.}~\bibnamefont {Predojevi\ifmmode~\acute{c}\else
  \'{c}\fi{}}}, \bibinfo {author} {\bibfnamefont {M.}~\bibnamefont
  {Koschorreck}},\ and\ \bibinfo {author} {\bibfnamefont {M.~W.}\ \bibnamefont
  {Mitchell}},\ }\bibfield  {title} {\bibinfo {title} {Squeezed-light optical
  magnetometry},\ }\href {https://doi.org/10.1103/PhysRevLett.105.053601}
  {\bibfield  {journal} {\bibinfo  {journal} {Phys. Rev. Lett.}\ }\textbf
  {\bibinfo {volume} {105}},\ \bibinfo {pages} {053601} (\bibinfo {year}
  {2010})}\BibitemShut {NoStop}%
\bibitem [{\citenamefont {Hosten}\ \emph {et~al.}(2016)\citenamefont {Hosten},
  \citenamefont {Engelsen}, \citenamefont {Krishnakumar},\ and\ \citenamefont
  {Kasevich}}]{Onur2016}%
  \BibitemOpen
  \bibfield  {author} {\bibinfo {author} {\bibfnamefont {O.}~\bibnamefont
  {Hosten}}, \bibinfo {author} {\bibfnamefont {N.~J.}\ \bibnamefont
  {Engelsen}}, \bibinfo {author} {\bibfnamefont {R.}~\bibnamefont
  {Krishnakumar}},\ and\ \bibinfo {author} {\bibfnamefont {M.~A.}\ \bibnamefont
  {Kasevich}},\ }\bibfield  {title} {\bibinfo {title} {Measurement noise 100
  times lower than the quantum-projection limit using entangled atoms},\ }\href
  {https://www.nature.com/articles/nature16176} {\bibfield  {journal} {\bibinfo
   {journal} {Nature}\ }\textbf {\bibinfo {volume} {529}},\ \bibinfo {pages}
  {505} (\bibinfo {year} {2016})}\BibitemShut {NoStop}%
\bibitem [{\citenamefont {Pezz\`e}\ \emph {et~al.}(2018)\citenamefont
  {Pezz\`e}, \citenamefont {Smerzi}, \citenamefont {Oberthaler}, \citenamefont
  {Schmied},\ and\ \citenamefont {Treutlein}}]{Pezz2018}%
  \BibitemOpen
  \bibfield  {author} {\bibinfo {author} {\bibfnamefont {L.}~\bibnamefont
  {Pezz\`e}}, \bibinfo {author} {\bibfnamefont {A.}~\bibnamefont {Smerzi}},
  \bibinfo {author} {\bibfnamefont {M.~K.}\ \bibnamefont {Oberthaler}},
  \bibinfo {author} {\bibfnamefont {R.}~\bibnamefont {Schmied}},\ and\ \bibinfo
  {author} {\bibfnamefont {P.}~\bibnamefont {Treutlein}},\ }\bibfield  {title}
  {\bibinfo {title} {Quantum metrology with nonclassical states of atomic
  ensembles},\ }\href {https://doi.org/10.1103/RevModPhys.90.035005} {\bibfield
   {journal} {\bibinfo  {journal} {Rev. Mod. Phys.}\ }\textbf {\bibinfo
  {volume} {90}},\ \bibinfo {pages} {035005} (\bibinfo {year}
  {2018})}\BibitemShut {NoStop}%
\bibitem [{\citenamefont {Chabuda}\ \emph {et~al.}(2020)\citenamefont
  {Chabuda}, \citenamefont {Dziarmaga}, \citenamefont {Osborne},\ and\
  \citenamefont {Demkowicz-Dobrzański}}]{Chabuda2020}%
  \BibitemOpen
  \bibfield  {author} {\bibinfo {author} {\bibfnamefont {K.}~\bibnamefont
  {Chabuda}}, \bibinfo {author} {\bibfnamefont {J.}~\bibnamefont {Dziarmaga}},
  \bibinfo {author} {\bibfnamefont {T.}~\bibnamefont {Osborne}},\ and\ \bibinfo
  {author} {\bibfnamefont {R.}~\bibnamefont {Demkowicz-Dobrzański}},\
  }\bibfield  {title} {\bibinfo {title} {Tensor-network approach for quantum
  metrology in many-body quantum systems},\ }\href
  {https://www.nature.com/articles/s41467-019-13735-9} {\bibfield  {journal}
  {\bibinfo  {journal} {Nat. Commun}\ }\textbf {\bibinfo {volume} {11}},\
  \bibinfo {pages} {250} (\bibinfo {year} {2020})}\BibitemShut {NoStop}%
\bibitem [{\citenamefont {Fadel}\ \emph {et~al.}(2018)\citenamefont {Fadel},
  \citenamefont {Zibold}, \citenamefont {Décamps},\ and\ \citenamefont
  {Treutlein}}]{Matteo2018}%
  \BibitemOpen
  \bibfield  {author} {\bibinfo {author} {\bibfnamefont {M.}~\bibnamefont
  {Fadel}}, \bibinfo {author} {\bibfnamefont {T.}~\bibnamefont {Zibold}},
  \bibinfo {author} {\bibfnamefont {B.}~\bibnamefont {Décamps}},\ and\
  \bibinfo {author} {\bibfnamefont {P.}~\bibnamefont {Treutlein}},\ }\bibfield
  {title} {\bibinfo {title} {Spatial entanglement patterns and
  einstein-podolsky-rosen steering in bose-einstein condensates},\ }\href
  {https://doi.org/10.1126/science.aao1850} {\bibfield  {journal} {\bibinfo
  {journal} {Science}\ }\textbf {\bibinfo {volume} {360}},\ \bibinfo {pages}
  {409} (\bibinfo {year} {2018})}\BibitemShut {NoStop}%
\bibitem [{\citenamefont {Tura}\ \emph {et~al.}(2014)\citenamefont {Tura},
  \citenamefont {Augusiak}, \citenamefont {Sainz}, \citenamefont {Vértesi},
  \citenamefont {Lewenstein},\ and\ \citenamefont {Acín}}]{Tura2014}%
  \BibitemOpen
  \bibfield  {author} {\bibinfo {author} {\bibfnamefont {J.}~\bibnamefont
  {Tura}}, \bibinfo {author} {\bibfnamefont {R.}~\bibnamefont {Augusiak}},
  \bibinfo {author} {\bibfnamefont {A.~B.}\ \bibnamefont {Sainz}}, \bibinfo
  {author} {\bibfnamefont {T.}~\bibnamefont {Vértesi}}, \bibinfo {author}
  {\bibfnamefont {M.}~\bibnamefont {Lewenstein}},\ and\ \bibinfo {author}
  {\bibfnamefont {A.}~\bibnamefont {Acín}},\ }\bibfield  {title} {\bibinfo
  {title} {Detecting nonlocality in many-body quantum states},\ }\href
  {https://doi.org/10.1126/science.1247715} {\bibfield  {journal} {\bibinfo
  {journal} {Science}\ }\textbf {\bibinfo {volume} {344}},\ \bibinfo {pages}
  {1256} (\bibinfo {year} {2014})}\BibitemShut {NoStop}%
\bibitem [{\citenamefont {Bao}\ \emph {et~al.}(2020)\citenamefont {Bao},
  \citenamefont {Duan}, \citenamefont {Jin}, \citenamefont {Lu}, \citenamefont
  {Li}, \citenamefont {Qu}, \citenamefont {Wang}, \citenamefont {Novikova},
  \citenamefont {Mikhailov}, \citenamefont {Zhao}, \citenamefont {Mølmer},
  \citenamefont {Shen},\ and\ \citenamefont {Xiao}}]{Han2020}%
  \BibitemOpen
  \bibfield  {author} {\bibinfo {author} {\bibfnamefont {H.}~\bibnamefont
  {Bao}}, \bibinfo {author} {\bibfnamefont {J.}~\bibnamefont {Duan}}, \bibinfo
  {author} {\bibfnamefont {S.}~\bibnamefont {Jin}}, \bibinfo {author}
  {\bibfnamefont {X.}~\bibnamefont {Lu}}, \bibinfo {author} {\bibfnamefont
  {P.}~\bibnamefont {Li}}, \bibinfo {author} {\bibfnamefont {W.}~\bibnamefont
  {Qu}}, \bibinfo {author} {\bibfnamefont {M.}~\bibnamefont {Wang}}, \bibinfo
  {author} {\bibfnamefont {I.}~\bibnamefont {Novikova}}, \bibinfo {author}
  {\bibfnamefont {E.~E.}\ \bibnamefont {Mikhailov}}, \bibinfo {author}
  {\bibfnamefont {K.-F.}\ \bibnamefont {Zhao}}, \bibinfo {author}
  {\bibfnamefont {K.}~\bibnamefont {Mølmer}}, \bibinfo {author} {\bibfnamefont
  {H.}~\bibnamefont {Shen}},\ and\ \bibinfo {author} {\bibfnamefont
  {Y.}~\bibnamefont {Xiao}},\ }\bibfield  {title} {\bibinfo {title} {Spin
  squeezing of 1011 atoms by prediction and retrodiction measurements},\ }\href
  {https://www.nature.com/articles/s41586-020-2243-7} {\bibfield  {journal}
  {\bibinfo  {journal} {Nature}\ }\textbf {\bibinfo {volume} {581}},\ \bibinfo
  {pages} {159} (\bibinfo {year} {2020})}\BibitemShut {NoStop}%
\bibitem [{\citenamefont {T\'oth}\ \emph {et~al.}(2009)\citenamefont {T\'oth},
  \citenamefont {Knapp}, \citenamefont {G\"uhne},\ and\ \citenamefont
  {Briegel}}]{Hans2009}%
  \BibitemOpen
  \bibfield  {author} {\bibinfo {author} {\bibfnamefont {G.}~\bibnamefont
  {T\'oth}}, \bibinfo {author} {\bibfnamefont {C.}~\bibnamefont {Knapp}},
  \bibinfo {author} {\bibfnamefont {O.}~\bibnamefont {G\"uhne}},\ and\ \bibinfo
  {author} {\bibfnamefont {H.~J.}\ \bibnamefont {Briegel}},\ }\bibfield
  {title} {\bibinfo {title} {Spin squeezing and entanglement},\ }\href
  {https://doi.org/10.1103/PhysRevA.79.042334} {\bibfield  {journal} {\bibinfo
  {journal} {Phys. Rev. A}\ }\textbf {\bibinfo {volume} {79}},\ \bibinfo
  {pages} {042334} (\bibinfo {year} {2009})}\BibitemShut {NoStop}%
\bibitem [{\citenamefont {Korbicz}\ \emph {et~al.}(2005)\citenamefont
  {Korbicz}, \citenamefont {Cirac},\ and\ \citenamefont
  {Lewenstein}}]{Korbicz2005}%
  \BibitemOpen
  \bibfield  {author} {\bibinfo {author} {\bibfnamefont {J.~K.}\ \bibnamefont
  {Korbicz}}, \bibinfo {author} {\bibfnamefont {J.~I.}\ \bibnamefont {Cirac}},\
  and\ \bibinfo {author} {\bibfnamefont {M.}~\bibnamefont {Lewenstein}},\
  }\bibfield  {title} {\bibinfo {title} {Spin squeezing inequalities and
  entanglement of $n$ qubit states},\ }\href
  {https://doi.org/10.1103/PhysRevLett.95.120502} {\bibfield  {journal}
  {\bibinfo  {journal} {Phys. Rev. Lett.}\ }\textbf {\bibinfo {volume} {95}},\
  \bibinfo {pages} {120502} (\bibinfo {year} {2005})}\BibitemShut {NoStop}%
\bibitem [{\citenamefont {Kuzmich}\ \emph {et~al.}(1997)\citenamefont
  {Kuzmich}, \citenamefont {M\o{}lmer},\ and\ \citenamefont
  {Polzik}}]{Kuzmich1997}%
  \BibitemOpen
  \bibfield  {author} {\bibinfo {author} {\bibfnamefont {A.}~\bibnamefont
  {Kuzmich}}, \bibinfo {author} {\bibfnamefont {K.}~\bibnamefont {M\o{}lmer}},\
  and\ \bibinfo {author} {\bibfnamefont {E.~S.}\ \bibnamefont {Polzik}},\
  }\bibfield  {title} {\bibinfo {title} {Spin squeezing in an ensemble of atoms
  illuminated with squeezed light},\ }\href
  {https://doi.org/10.1103/PhysRevLett.79.4782} {\bibfield  {journal} {\bibinfo
   {journal} {Phys. Rev. Lett.}\ }\textbf {\bibinfo {volume} {79}},\ \bibinfo
  {pages} {4782} (\bibinfo {year} {1997})}\BibitemShut {NoStop}%
\bibitem [{\citenamefont {Hald}\ \emph {et~al.}(1999)\citenamefont {Hald},
  \citenamefont {S\o{}rensen}, \citenamefont {Schori},\ and\ \citenamefont
  {Polzik}}]{Hald1999}%
  \BibitemOpen
  \bibfield  {author} {\bibinfo {author} {\bibfnamefont {J.}~\bibnamefont
  {Hald}}, \bibinfo {author} {\bibfnamefont {J.~L.}\ \bibnamefont
  {S\o{}rensen}}, \bibinfo {author} {\bibfnamefont {C.}~\bibnamefont
  {Schori}},\ and\ \bibinfo {author} {\bibfnamefont {E.~S.}\ \bibnamefont
  {Polzik}},\ }\bibfield  {title} {\bibinfo {title} {Spin squeezed atoms: A
  macroscopic entangled ensemble created by light},\ }\href
  {https://doi.org/10.1103/PhysRevLett.83.1319} {\bibfield  {journal} {\bibinfo
   {journal} {Phys. Rev. Lett.}\ }\textbf {\bibinfo {volume} {83}},\ \bibinfo
  {pages} {1319} (\bibinfo {year} {1999})}\BibitemShut {NoStop}%
\bibitem [{\citenamefont {Vernac}\ \emph {et~al.}(2000)\citenamefont {Vernac},
  \citenamefont {Pinard},\ and\ \citenamefont {Giacobino}}]{Vernac2000}%
  \BibitemOpen
  \bibfield  {author} {\bibinfo {author} {\bibfnamefont {L.}~\bibnamefont
  {Vernac}}, \bibinfo {author} {\bibfnamefont {M.}~\bibnamefont {Pinard}},\
  and\ \bibinfo {author} {\bibfnamefont {E.}~\bibnamefont {Giacobino}},\
  }\bibfield  {title} {\bibinfo {title} {Spin squeezing in two-level systems},\
  }\href {https://doi.org/10.1103/PhysRevA.62.063812} {\bibfield  {journal}
  {\bibinfo  {journal} {Phys. Rev. A}\ }\textbf {\bibinfo {volume} {62}},\
  \bibinfo {pages} {063812} (\bibinfo {year} {2000})}\BibitemShut {NoStop}%
\bibitem [{\citenamefont {Fleischhauer}\ and\ \citenamefont
  {Gong}(2002)}]{Fleischhauer2002}%
  \BibitemOpen
  \bibfield  {author} {\bibinfo {author} {\bibfnamefont {M.}~\bibnamefont
  {Fleischhauer}}\ and\ \bibinfo {author} {\bibfnamefont {S.}~\bibnamefont
  {Gong}},\ }\bibfield  {title} {\bibinfo {title} {Stationary source of
  nonclassical or entangled atoms},\ }\href
  {https://doi.org/10.1103/PhysRevLett.88.070404} {\bibfield  {journal}
  {\bibinfo  {journal} {Phys. Rev. Lett.}\ }\textbf {\bibinfo {volume} {88}},\
  \bibinfo {pages} {070404} (\bibinfo {year} {2002})}\BibitemShut {NoStop}%
\bibitem [{\citenamefont {Kuzmich}\ \emph {et~al.}(1999)\citenamefont
  {Kuzmich}, \citenamefont {Mandel}, \citenamefont {Janis}, \citenamefont
  {Young}, \citenamefont {Ejnisman},\ and\ \citenamefont
  {Bigelow}}]{Kuzmich1999}%
  \BibitemOpen
  \bibfield  {author} {\bibinfo {author} {\bibfnamefont {A.}~\bibnamefont
  {Kuzmich}}, \bibinfo {author} {\bibfnamefont {L.}~\bibnamefont {Mandel}},
  \bibinfo {author} {\bibfnamefont {J.}~\bibnamefont {Janis}}, \bibinfo
  {author} {\bibfnamefont {Y.~E.}\ \bibnamefont {Young}}, \bibinfo {author}
  {\bibfnamefont {R.}~\bibnamefont {Ejnisman}},\ and\ \bibinfo {author}
  {\bibfnamefont {N.~P.}\ \bibnamefont {Bigelow}},\ }\bibfield  {title}
  {\bibinfo {title} {Quantum nondemolition measurements of collective atomic
  spin},\ }\href {https://doi.org/10.1103/PhysRevA.60.2346} {\bibfield
  {journal} {\bibinfo  {journal} {Phys. Rev. A}\ }\textbf {\bibinfo {volume}
  {60}},\ \bibinfo {pages} {2346} (\bibinfo {year} {1999})}\BibitemShut
  {NoStop}%
\bibitem [{\citenamefont {Kuzmich}\ \emph {et~al.}(2000)\citenamefont
  {Kuzmich}, \citenamefont {Mandel},\ and\ \citenamefont
  {Bigelow}}]{Kuzmich2000}%
  \BibitemOpen
  \bibfield  {author} {\bibinfo {author} {\bibfnamefont {A.}~\bibnamefont
  {Kuzmich}}, \bibinfo {author} {\bibfnamefont {L.}~\bibnamefont {Mandel}},\
  and\ \bibinfo {author} {\bibfnamefont {N.~P.}\ \bibnamefont {Bigelow}},\
  }\bibfield  {title} {\bibinfo {title} {Generation of spin squeezing via
  continuous quantum nondemolition measurement},\ }\href
  {https://doi.org/10.1103/PhysRevLett.85.1594} {\bibfield  {journal} {\bibinfo
   {journal} {Phys. Rev. Lett.}\ }\textbf {\bibinfo {volume} {85}},\ \bibinfo
  {pages} {1594} (\bibinfo {year} {2000})}\BibitemShut {NoStop}%
\bibitem [{\citenamefont {Inoue}\ \emph {et~al.}(2013)\citenamefont {Inoue},
  \citenamefont {Tanaka}, \citenamefont {Namiki}, \citenamefont {Sagawa},\ and\
  \citenamefont {Takahashi}}]{Inoue2013}%
  \BibitemOpen
  \bibfield  {author} {\bibinfo {author} {\bibfnamefont {R.}~\bibnamefont
  {Inoue}}, \bibinfo {author} {\bibfnamefont {S.-I.-R.}\ \bibnamefont
  {Tanaka}}, \bibinfo {author} {\bibfnamefont {R.}~\bibnamefont {Namiki}},
  \bibinfo {author} {\bibfnamefont {T.}~\bibnamefont {Sagawa}},\ and\ \bibinfo
  {author} {\bibfnamefont {Y.}~\bibnamefont {Takahashi}},\ }\bibfield  {title}
  {\bibinfo {title} {Unconditional quantum-noise suppression via
  measurement-based quantum feedback},\ }\href
  {https://doi.org/10.1103/PhysRevLett.110.163602} {\bibfield  {journal}
  {\bibinfo  {journal} {Phys. Rev. Lett.}\ }\textbf {\bibinfo {volume} {110}},\
  \bibinfo {pages} {163602} (\bibinfo {year} {2013})}\BibitemShut {NoStop}%
\bibitem [{\citenamefont {Rossi}\ \emph {et~al.}(2020)\citenamefont {Rossi},
  \citenamefont {Albarelli}, \citenamefont {Tamascelli},\ and\ \citenamefont
  {Genoni}}]{Rossi2020}%
  \BibitemOpen
  \bibfield  {author} {\bibinfo {author} {\bibfnamefont {M.~A.~C.}\
  \bibnamefont {Rossi}}, \bibinfo {author} {\bibfnamefont {F.}~\bibnamefont
  {Albarelli}}, \bibinfo {author} {\bibfnamefont {D.}~\bibnamefont
  {Tamascelli}},\ and\ \bibinfo {author} {\bibfnamefont {M.~G.}\ \bibnamefont
  {Genoni}},\ }\bibfield  {title} {\bibinfo {title} {Noisy quantum metrology
  enhanced by continuous nondemolition measurement},\ }\href
  {https://doi.org/10.1103/PhysRevLett.125.200505} {\bibfield  {journal}
  {\bibinfo  {journal} {Phys. Rev. Lett.}\ }\textbf {\bibinfo {volume} {125}},\
  \bibinfo {pages} {200505} (\bibinfo {year} {2020})}\BibitemShut {NoStop}%
\bibitem [{\citenamefont {Liu}\ \emph {et~al.}(2011)\citenamefont {Liu},
  \citenamefont {Xu}, \citenamefont {Jin},\ and\ \citenamefont
  {You}}]{Liu2011}%
  \BibitemOpen
  \bibfield  {author} {\bibinfo {author} {\bibfnamefont {Y.~C.}\ \bibnamefont
  {Liu}}, \bibinfo {author} {\bibfnamefont {Z.~F.}\ \bibnamefont {Xu}},
  \bibinfo {author} {\bibfnamefont {G.~R.}\ \bibnamefont {Jin}},\ and\ \bibinfo
  {author} {\bibfnamefont {L.}~\bibnamefont {You}},\ }\bibfield  {title}
  {\bibinfo {title} {Spin squeezing: Transforming one-axis twisting into
  two-axis twisting},\ }\href {https://doi.org/10.1103/PhysRevLett.107.013601}
  {\bibfield  {journal} {\bibinfo  {journal} {Phys. Rev. Lett.}\ }\textbf
  {\bibinfo {volume} {107}},\ \bibinfo {pages} {013601} (\bibinfo {year}
  {2011})}\BibitemShut {NoStop}%
\bibitem [{\citenamefont {Zhang}\ \emph
  {et~al.}(2017{\natexlab{a}})\citenamefont {Zhang}, \citenamefont {Zhou},
  \citenamefont {Zhou}, \citenamefont {Guo},\ and\ \citenamefont
  {Zhou}}]{Zhang2017}%
  \BibitemOpen
  \bibfield  {author} {\bibinfo {author} {\bibfnamefont {Y.-C.}\ \bibnamefont
  {Zhang}}, \bibinfo {author} {\bibfnamefont {X.-F.}\ \bibnamefont {Zhou}},
  \bibinfo {author} {\bibfnamefont {X.}~\bibnamefont {Zhou}}, \bibinfo {author}
  {\bibfnamefont {G.-C.}\ \bibnamefont {Guo}},\ and\ \bibinfo {author}
  {\bibfnamefont {Z.-W.}\ \bibnamefont {Zhou}},\ }\bibfield  {title} {\bibinfo
  {title} {Cavity-assisted single-mode and two-mode spin-squeezed states via
  phase-locked atom-photon coupling},\ }\href
  {https://doi.org/10.1103/PhysRevLett.118.083604} {\bibfield  {journal}
  {\bibinfo  {journal} {Phys. Rev. Lett.}\ }\textbf {\bibinfo {volume} {118}},\
  \bibinfo {pages} {083604} (\bibinfo {year} {2017}{\natexlab{a}})}\BibitemShut
  {NoStop}%
\bibitem [{\citenamefont {Groszkowski}\ \emph {et~al.}(2020)\citenamefont
  {Groszkowski}, \citenamefont {Lau}, \citenamefont {Leroux}, \citenamefont
  {Govia},\ and\ \citenamefont {Clerk}}]{Groszkowski2020}%
  \BibitemOpen
  \bibfield  {author} {\bibinfo {author} {\bibfnamefont {P.}~\bibnamefont
  {Groszkowski}}, \bibinfo {author} {\bibfnamefont {H.-K.}\ \bibnamefont
  {Lau}}, \bibinfo {author} {\bibfnamefont {C.}~\bibnamefont {Leroux}},
  \bibinfo {author} {\bibfnamefont {L.~C.~G.}\ \bibnamefont {Govia}},\ and\
  \bibinfo {author} {\bibfnamefont {A.~A.}\ \bibnamefont {Clerk}},\ }\bibfield
  {title} {\bibinfo {title} {Heisenberg-limited spin squeezing via bosonic
  parametric driving},\ }\href {https://doi.org/10.1103/PhysRevLett.125.203601}
  {\bibfield  {journal} {\bibinfo  {journal} {Phys. Rev. Lett.}\ }\textbf
  {\bibinfo {volume} {125}},\ \bibinfo {pages} {203601} (\bibinfo {year}
  {2020})}\BibitemShut {NoStop}%
\bibitem [{\citenamefont {Huang}\ \emph {et~al.}(2021)\citenamefont {Huang},
  \citenamefont {Chen}, \citenamefont {Li}, \citenamefont {Li}, \citenamefont
  {Lü},\ and\ \citenamefont {Liu}}]{Huang2021}%
  \BibitemOpen
  \bibfield  {author} {\bibinfo {author} {\bibfnamefont {L.-G.}\ \bibnamefont
  {Huang}}, \bibinfo {author} {\bibfnamefont {F.}~\bibnamefont {Chen}},
  \bibinfo {author} {\bibfnamefont {X.}~\bibnamefont {Li}}, \bibinfo {author}
  {\bibfnamefont {Y.}~\bibnamefont {Li}}, \bibinfo {author} {\bibfnamefont
  {R.}~\bibnamefont {Lü}},\ and\ \bibinfo {author} {\bibfnamefont {Y.-C.}\
  \bibnamefont {Liu}},\ }\bibfield  {title} {\bibinfo {title} {Dynamic
  synthesis of heisenberg-limited spin squeezing},\ }\href
  {https://www.nature.com/articles/s41534-021-00505-z#citeas} {\bibfield
  {journal} {\bibinfo  {journal} {npj Quantum Information}\ }\textbf {\bibinfo
  {volume} {7}},\ \bibinfo {pages} {168} (\bibinfo {year} {2021})}\BibitemShut
  {NoStop}%
\bibitem [{\citenamefont {Zhang}\ \emph {et~al.}(2014)\citenamefont {Zhang},
  \citenamefont {Zhou}, \citenamefont {Guo},\ and\ \citenamefont
  {Zhou}}]{Zhang2014}%
  \BibitemOpen
  \bibfield  {author} {\bibinfo {author} {\bibfnamefont {J.-Y.}\ \bibnamefont
  {Zhang}}, \bibinfo {author} {\bibfnamefont {X.-F.}\ \bibnamefont {Zhou}},
  \bibinfo {author} {\bibfnamefont {G.-C.}\ \bibnamefont {Guo}},\ and\ \bibinfo
  {author} {\bibfnamefont {Z.-W.}\ \bibnamefont {Zhou}},\ }\bibfield  {title}
  {\bibinfo {title} {Dynamical spin squeezing via a higher-order trotter-suzuki
  approximation},\ }\href {https://doi.org/10.1103/PhysRevA.90.013604}
  {\bibfield  {journal} {\bibinfo  {journal} {Phys. Rev. A}\ }\textbf {\bibinfo
  {volume} {90}},\ \bibinfo {pages} {013604} (\bibinfo {year}
  {2014})}\BibitemShut {NoStop}%
\bibitem [{\citenamefont {Huang}\ \emph {et~al.}(2015)\citenamefont {Huang},
  \citenamefont {Zhang}, \citenamefont {Zou}, \citenamefont {Zou},\ and\
  \citenamefont {Guo}}]{Huang2015}%
  \BibitemOpen
  \bibfield  {author} {\bibinfo {author} {\bibfnamefont {W.}~\bibnamefont
  {Huang}}, \bibinfo {author} {\bibfnamefont {Y.-L.}\ \bibnamefont {Zhang}},
  \bibinfo {author} {\bibfnamefont {C.-L.}\ \bibnamefont {Zou}}, \bibinfo
  {author} {\bibfnamefont {X.-B.}\ \bibnamefont {Zou}},\ and\ \bibinfo {author}
  {\bibfnamefont {G.-C.}\ \bibnamefont {Guo}},\ }\bibfield  {title} {\bibinfo
  {title} {Two-axis spin squeezing of two-component bose-einstein condensates
  via continuous driving},\ }\href {https://doi.org/10.1103/PhysRevA.91.043642}
  {\bibfield  {journal} {\bibinfo  {journal} {Phys. Rev. A}\ }\textbf {\bibinfo
  {volume} {91}},\ \bibinfo {pages} {043642} (\bibinfo {year}
  {2015})}\BibitemShut {NoStop}%
\bibitem [{\citenamefont {Bai}\ and\ \citenamefont {An}(2021)}]{Bai2021}%
  \BibitemOpen
  \bibfield  {author} {\bibinfo {author} {\bibfnamefont {S.-Y.}\ \bibnamefont
  {Bai}}\ and\ \bibinfo {author} {\bibfnamefont {J.-H.}\ \bibnamefont {An}},\
  }\bibfield  {title} {\bibinfo {title} {Generating stable spin squeezing by
  squeezed-reservoir engineering},\ }\href
  {https://doi.org/10.1103/PhysRevLett.127.083602} {\bibfield  {journal}
  {\bibinfo  {journal} {Phys. Rev. Lett.}\ }\textbf {\bibinfo {volume} {127}},\
  \bibinfo {pages} {083602} (\bibinfo {year} {2021})}\BibitemShut {NoStop}%
\bibitem [{\citenamefont {Helmerson}\ and\ \citenamefont
  {You}(2001)}]{Helmerson2001}%
  \BibitemOpen
  \bibfield  {author} {\bibinfo {author} {\bibfnamefont {K.}~\bibnamefont
  {Helmerson}}\ and\ \bibinfo {author} {\bibfnamefont {L.}~\bibnamefont
  {You}},\ }\bibfield  {title} {\bibinfo {title} {Creating massive entanglement
  of bose-einstein condensed atoms},\ }\href
  {https://doi.org/10.1103/PhysRevLett.87.170402} {\bibfield  {journal}
  {\bibinfo  {journal} {Phys. Rev. Lett.}\ }\textbf {\bibinfo {volume} {87}},\
  \bibinfo {pages} {170402} (\bibinfo {year} {2001})}\BibitemShut {NoStop}%
\bibitem [{\citenamefont {Jin}\ \emph {et~al.}(2021)\citenamefont {Jin},
  \citenamefont {Bao}, \citenamefont {Duan}, \citenamefont {Lu}, \citenamefont
  {Wang}, \citenamefont {Zhao}, \citenamefont {Shen},\ and\ \citenamefont
  {Xiao}}]{Shenchao2021}%
  \BibitemOpen
  \bibfield  {author} {\bibinfo {author} {\bibfnamefont {S.}~\bibnamefont
  {Jin}}, \bibinfo {author} {\bibfnamefont {H.}~\bibnamefont {Bao}}, \bibinfo
  {author} {\bibfnamefont {J.}~\bibnamefont {Duan}}, \bibinfo {author}
  {\bibfnamefont {X.}~\bibnamefont {Lu}}, \bibinfo {author} {\bibfnamefont
  {M.}~\bibnamefont {Wang}}, \bibinfo {author} {\bibfnamefont {K.-F.}\
  \bibnamefont {Zhao}}, \bibinfo {author} {\bibfnamefont {H.}~\bibnamefont
  {Shen}},\ and\ \bibinfo {author} {\bibfnamefont {Y.}~\bibnamefont {Xiao}},\
  }\bibfield  {title} {\bibinfo {title} {Adiabaticity in state preparation for
  spin squeezing of large atom ensembles},\ }\href
  {https://www.researching.cn/articles/OJ22e7f554c8b7d337} {\bibfield
  {journal} {\bibinfo  {journal} {Photon. Res.}\ }\textbf {\bibinfo {volume}
  {9}},\ \bibinfo {pages} {2296} (\bibinfo {year} {2021})}\BibitemShut
  {NoStop}%
\bibitem [{\citenamefont {Zhang}\ \emph {et~al.}(2003)\citenamefont {Zhang},
  \citenamefont {Helmerson},\ and\ \citenamefont {You}}]{Zhang2003}%
  \BibitemOpen
  \bibfield  {author} {\bibinfo {author} {\bibfnamefont {M.}~\bibnamefont
  {Zhang}}, \bibinfo {author} {\bibfnamefont {K.}~\bibnamefont {Helmerson}},\
  and\ \bibinfo {author} {\bibfnamefont {L.}~\bibnamefont {You}},\ }\bibfield
  {title} {\bibinfo {title} {Entanglement and spin squeezing of
  bose-einstein-condensed atoms},\ }\href
  {https://doi.org/10.1103/PhysRevA.68.043622} {\bibfield  {journal} {\bibinfo
  {journal} {Phys. Rev. A}\ }\textbf {\bibinfo {volume} {68}},\ \bibinfo
  {pages} {043622} (\bibinfo {year} {2003})}\BibitemShut {NoStop}%
\bibitem [{\citenamefont {Bohnet}\ \emph {et~al.}(2016)\citenamefont {Bohnet},
  \citenamefont {Sawyer}, \citenamefont {Britton}, \citenamefont {Wall},
  \citenamefont {Rey}, \citenamefont {Foss-Feig},\ and\ \citenamefont
  {Bollinger}}]{Justin2016}%
  \BibitemOpen
  \bibfield  {author} {\bibinfo {author} {\bibfnamefont {J.~G.}\ \bibnamefont
  {Bohnet}}, \bibinfo {author} {\bibfnamefont {B.~C.}\ \bibnamefont {Sawyer}},
  \bibinfo {author} {\bibfnamefont {J.~W.}\ \bibnamefont {Britton}}, \bibinfo
  {author} {\bibfnamefont {M.~L.}\ \bibnamefont {Wall}}, \bibinfo {author}
  {\bibfnamefont {A.~M.}\ \bibnamefont {Rey}}, \bibinfo {author} {\bibfnamefont
  {M.}~\bibnamefont {Foss-Feig}},\ and\ \bibinfo {author} {\bibfnamefont
  {J.~J.}\ \bibnamefont {Bollinger}},\ }\bibfield  {title} {\bibinfo {title}
  {Quantum spin dynamics and entanglement generation with hundreds of trapped
  ions},\ }\href {https://doi.org/10.1126/science.aad9958} {\bibfield
  {journal} {\bibinfo  {journal} {Science}\ }\textbf {\bibinfo {volume}
  {352}},\ \bibinfo {pages} {1297} (\bibinfo {year} {2016})}\BibitemShut
  {NoStop}%
\bibitem [{\citenamefont {Wu}\ \emph {et~al.}(2015)\citenamefont {Wu},
  \citenamefont {Tey},\ and\ \citenamefont {You}}]{Wu2015}%
  \BibitemOpen
  \bibfield  {author} {\bibinfo {author} {\bibfnamefont {L.-N.}\ \bibnamefont
  {Wu}}, \bibinfo {author} {\bibfnamefont {M.~K.}\ \bibnamefont {Tey}},\ and\
  \bibinfo {author} {\bibfnamefont {L.}~\bibnamefont {You}},\ }\bibfield
  {title} {\bibinfo {title} {Persistent atomic spin squeezing at the heisenberg
  limit},\ }\href {https://doi.org/10.1103/PhysRevA.92.063610} {\bibfield
  {journal} {\bibinfo  {journal} {Phys. Rev. A}\ }\textbf {\bibinfo {volume}
  {92}},\ \bibinfo {pages} {063610} (\bibinfo {year} {2015})}\BibitemShut
  {NoStop}%
\bibitem [{\citenamefont {Wang}\ \emph
  {et~al.}(2017{\natexlab{a}})\citenamefont {Wang}, \citenamefont {Qu},
  \citenamefont {Li}, \citenamefont {Bao}, \citenamefont
  {Vuleti\ifmmode~\acute{c}\else \'{c}\fi{}},\ and\ \citenamefont
  {Xiao}}]{Wang2017}%
  \BibitemOpen
  \bibfield  {author} {\bibinfo {author} {\bibfnamefont {M.}~\bibnamefont
  {Wang}}, \bibinfo {author} {\bibfnamefont {W.}~\bibnamefont {Qu}}, \bibinfo
  {author} {\bibfnamefont {P.}~\bibnamefont {Li}}, \bibinfo {author}
  {\bibfnamefont {H.}~\bibnamefont {Bao}}, \bibinfo {author} {\bibfnamefont
  {V.}~\bibnamefont {Vuleti\ifmmode~\acute{c}\else \'{c}\fi{}}},\ and\ \bibinfo
  {author} {\bibfnamefont {Y.}~\bibnamefont {Xiao}},\ }\bibfield  {title}
  {\bibinfo {title} {Two-axis-twisting spin squeezing by multipass quantum
  erasure},\ }\href {https://doi.org/10.1103/PhysRevA.96.013823} {\bibfield
  {journal} {\bibinfo  {journal} {Phys. Rev. A}\ }\textbf {\bibinfo {volume}
  {96}},\ \bibinfo {pages} {013823} (\bibinfo {year}
  {2017}{\natexlab{a}})}\BibitemShut {NoStop}%
\bibitem [{\citenamefont {Kitzinger}\ \emph {et~al.}(2020)\citenamefont
  {Kitzinger}, \citenamefont {Chaudhary}, \citenamefont {Kondappan},
  \citenamefont {Ivannikov},\ and\ \citenamefont {Byrnes}}]{Kitzinger2020}%
  \BibitemOpen
  \bibfield  {author} {\bibinfo {author} {\bibfnamefont {J.}~\bibnamefont
  {Kitzinger}}, \bibinfo {author} {\bibfnamefont {M.}~\bibnamefont
  {Chaudhary}}, \bibinfo {author} {\bibfnamefont {M.}~\bibnamefont
  {Kondappan}}, \bibinfo {author} {\bibfnamefont {V.}~\bibnamefont
  {Ivannikov}},\ and\ \bibinfo {author} {\bibfnamefont {T.}~\bibnamefont
  {Byrnes}},\ }\bibfield  {title} {\bibinfo {title} {Two-axis two-spin squeezed
  states},\ }\href {https://doi.org/10.1103/PhysRevResearch.2.033504}
  {\bibfield  {journal} {\bibinfo  {journal} {Phys. Rev. Res.}\ }\textbf
  {\bibinfo {volume} {2}},\ \bibinfo {pages} {033504} (\bibinfo {year}
  {2020})}\BibitemShut {NoStop}%
\bibitem [{\citenamefont {Hern\'andez~Yanes}\ \emph {et~al.}(2022)\citenamefont
  {Hern\'andez~Yanes}, \citenamefont {P\l{}odzie\ifmmode~\acute{n}\else
  \'{n}\fi{}}, \citenamefont {Mackoit Sinkevi\ifmmode \check{c}\else
  \v{c}\fi{}ien\ifmmode~\dot{e}\else \.{e}\fi{}}, \citenamefont
  {\ifmmode~\check{Z}\else \v{Z}\fi{}labys}, \citenamefont
  {Juzeli\ifmmode~\bar{u}\else \={u}\fi{}nas},\ and\ \citenamefont
  {Witkowska}}]{Witkowska2022}%
  \BibitemOpen
  \bibfield  {author} {\bibinfo {author} {\bibfnamefont {T.}~\bibnamefont
  {Hern\'andez~Yanes}}, \bibinfo {author} {\bibfnamefont {M.}~\bibnamefont
  {P\l{}odzie\ifmmode~\acute{n}\else \'{n}\fi{}}}, \bibinfo {author}
  {\bibfnamefont {M.}~\bibnamefont {Mackoit Sinkevi\ifmmode \check{c}\else
  \v{c}\fi{}ien\ifmmode~\dot{e}\else \.{e}\fi{}}}, \bibinfo {author}
  {\bibfnamefont {G.}~\bibnamefont {\ifmmode~\check{Z}\else \v{Z}\fi{}labys}},
  \bibinfo {author} {\bibfnamefont {G.}~\bibnamefont
  {Juzeli\ifmmode~\bar{u}\else \={u}\fi{}nas}},\ and\ \bibinfo {author}
  {\bibfnamefont {E.}~\bibnamefont {Witkowska}},\ }\bibfield  {title} {\bibinfo
  {title} {One- and two-axis squeezing via laser coupling in an atomic
  fermi-hubbard model},\ }\href
  {https://doi.org/10.1103/PhysRevLett.129.090403} {\bibfield  {journal}
  {\bibinfo  {journal} {Phys. Rev. Lett.}\ }\textbf {\bibinfo {volume} {129}},\
  \bibinfo {pages} {090403} (\bibinfo {year} {2022})}\BibitemShut {NoStop}%
\bibitem [{\citenamefont {Huang}\ \emph {et~al.}(2023)\citenamefont {Huang},
  \citenamefont {Zhang}, \citenamefont {Wang}, \citenamefont {Hua},
  \citenamefont {Tang},\ and\ \citenamefont {Liu}}]{Huang2023}%
  \BibitemOpen
  \bibfield  {author} {\bibinfo {author} {\bibfnamefont {L.-G.}\ \bibnamefont
  {Huang}}, \bibinfo {author} {\bibfnamefont {X.}~\bibnamefont {Zhang}},
  \bibinfo {author} {\bibfnamefont {Y.}~\bibnamefont {Wang}}, \bibinfo {author}
  {\bibfnamefont {Z.}~\bibnamefont {Hua}}, \bibinfo {author} {\bibfnamefont
  {Y.}~\bibnamefont {Tang}},\ and\ \bibinfo {author} {\bibfnamefont {Y.-C.}\
  \bibnamefont {Liu}},\ }\bibfield  {title} {\bibinfo {title}
  {Heisenberg-limited spin squeezing in coupled spin systems},\ }\href
  {https://doi.org/10.1103/PhysRevA.107.042613} {\bibfield  {journal} {\bibinfo
   {journal} {Phys. Rev. A}\ }\textbf {\bibinfo {volume} {107}},\ \bibinfo
  {pages} {042613} (\bibinfo {year} {2023})}\BibitemShut {NoStop}%
\bibitem [{\citenamefont {Sipahigil}\ \emph {et~al.}(2014)\citenamefont
  {Sipahigil}, \citenamefont {Jahnke}, \citenamefont {Rogers}, \citenamefont
  {Teraji}, \citenamefont {Isoya}, \citenamefont {Zibrov}, \citenamefont
  {Jelezko},\ and\ \citenamefont {Lukin}}]{Sipahigil2014}%
  \BibitemOpen
  \bibfield  {author} {\bibinfo {author} {\bibfnamefont {A.}~\bibnamefont
  {Sipahigil}}, \bibinfo {author} {\bibfnamefont {K.~D.}\ \bibnamefont
  {Jahnke}}, \bibinfo {author} {\bibfnamefont {L.~J.}\ \bibnamefont {Rogers}},
  \bibinfo {author} {\bibfnamefont {T.}~\bibnamefont {Teraji}}, \bibinfo
  {author} {\bibfnamefont {J.}~\bibnamefont {Isoya}}, \bibinfo {author}
  {\bibfnamefont {A.~S.}\ \bibnamefont {Zibrov}}, \bibinfo {author}
  {\bibfnamefont {F.}~\bibnamefont {Jelezko}},\ and\ \bibinfo {author}
  {\bibfnamefont {M.~D.}\ \bibnamefont {Lukin}},\ }\bibfield  {title} {\bibinfo
  {title} {Indistinguishable photons from separated silicon-vacancy centers in
  diamond},\ }\href {https://doi.org/10.1103/PhysRevLett.113.113602} {\bibfield
   {journal} {\bibinfo  {journal} {Phys. Rev. Lett.}\ }\textbf {\bibinfo
  {volume} {113}},\ \bibinfo {pages} {113602} (\bibinfo {year}
  {2014})}\BibitemShut {NoStop}%
\bibitem [{\citenamefont {Li}\ \emph {et~al.}(2016)\citenamefont {Li},
  \citenamefont {Xiang}, \citenamefont {Rabl},\ and\ \citenamefont
  {Nori}}]{Li2016}%
  \BibitemOpen
  \bibfield  {author} {\bibinfo {author} {\bibfnamefont {P.-B.}\ \bibnamefont
  {Li}}, \bibinfo {author} {\bibfnamefont {Z.-L.}\ \bibnamefont {Xiang}},
  \bibinfo {author} {\bibfnamefont {P.}~\bibnamefont {Rabl}},\ and\ \bibinfo
  {author} {\bibfnamefont {F.}~\bibnamefont {Nori}},\ }\bibfield  {title}
  {\bibinfo {title} {Hybrid quantum device with nitrogen-vacancy centers in
  diamond coupled to carbon nanotubes},\ }\href
  {https://doi.org/10.1103/PhysRevLett.117.015502} {\bibfield  {journal}
  {\bibinfo  {journal} {Phys. Rev. Lett.}\ }\textbf {\bibinfo {volume} {117}},\
  \bibinfo {pages} {015502} (\bibinfo {year} {2016})}\BibitemShut {NoStop}%
\bibitem [{\citenamefont {Golter}\ \emph {et~al.}(2016)\citenamefont {Golter},
  \citenamefont {Oo}, \citenamefont {Amezcua}, \citenamefont {Lekavicius},
  \citenamefont {Stewart},\ and\ \citenamefont {Wang}}]{Golter2016}%
  \BibitemOpen
  \bibfield  {author} {\bibinfo {author} {\bibfnamefont {D.~A.}\ \bibnamefont
  {Golter}}, \bibinfo {author} {\bibfnamefont {T.}~\bibnamefont {Oo}}, \bibinfo
  {author} {\bibfnamefont {M.}~\bibnamefont {Amezcua}}, \bibinfo {author}
  {\bibfnamefont {I.}~\bibnamefont {Lekavicius}}, \bibinfo {author}
  {\bibfnamefont {K.~A.}\ \bibnamefont {Stewart}},\ and\ \bibinfo {author}
  {\bibfnamefont {H.}~\bibnamefont {Wang}},\ }\bibfield  {title} {\bibinfo
  {title} {Coupling a surface acoustic wave to an electron spin in diamond via
  a dark state},\ }\href {https://doi.org/10.1103/PhysRevX.6.041060} {\bibfield
   {journal} {\bibinfo  {journal} {Phys. Rev. X}\ }\textbf {\bibinfo {volume}
  {6}},\ \bibinfo {pages} {041060} (\bibinfo {year} {2016})}\BibitemShut
  {NoStop}%
\bibitem [{\citenamefont {Song}\ \emph {et~al.}(2017)\citenamefont {Song},
  \citenamefont {Yang}, \citenamefont {An},\ and\ \citenamefont
  {Feng}}]{Song2017}%
  \BibitemOpen
  \bibfield  {author} {\bibinfo {author} {\bibfnamefont {W.}~\bibnamefont
  {Song}}, \bibinfo {author} {\bibfnamefont {W.}~\bibnamefont {Yang}}, \bibinfo
  {author} {\bibfnamefont {J.}~\bibnamefont {An}},\ and\ \bibinfo {author}
  {\bibfnamefont {M.}~\bibnamefont {Feng}},\ }\bibfield  {title} {\bibinfo
  {title} {Dissipation-assisted spin squeezing of nitrogen-vacancy centers
  coupled to a rectangular hollow metallic waveguide},\ }\href
  {https://doi.org/10.1364/OE.25.019226} {\bibfield  {journal} {\bibinfo
  {journal} {Opt. Express}\ }\textbf {\bibinfo {volume} {25}},\ \bibinfo
  {pages} {19226} (\bibinfo {year} {2017})}\BibitemShut {NoStop}%
\bibitem [{\citenamefont {Li}\ \emph {et~al.}(2020{\natexlab{a}})\citenamefont
  {Li}, \citenamefont {Zhou}, \citenamefont {Gao},\ and\ \citenamefont
  {Nori}}]{Lipengbo2020}%
  \BibitemOpen
  \bibfield  {author} {\bibinfo {author} {\bibfnamefont {P.-B.}\ \bibnamefont
  {Li}}, \bibinfo {author} {\bibfnamefont {Y.}~\bibnamefont {Zhou}}, \bibinfo
  {author} {\bibfnamefont {W.-B.}\ \bibnamefont {Gao}},\ and\ \bibinfo {author}
  {\bibfnamefont {F.}~\bibnamefont {Nori}},\ }\bibfield  {title} {\bibinfo
  {title} {Enhancing spin-phonon and spin-spin interactions using linear
  resources in a hybrid quantum system},\ }\href
  {https://doi.org/10.1103/PhysRevLett.125.153602} {\bibfield  {journal}
  {\bibinfo  {journal} {Phys. Rev. Lett.}\ }\textbf {\bibinfo {volume} {125}},\
  \bibinfo {pages} {153602} (\bibinfo {year} {2020}{\natexlab{a}})}\BibitemShut
  {NoStop}%
\bibitem [{\citenamefont {Zhou}\ \emph {et~al.}(2022)\citenamefont {Zhou},
  \citenamefont {Hu}, \citenamefont {L\"{u}}, \citenamefont {Li}, \citenamefont
  {Huang}, \citenamefont {Xiong},\ and\ \citenamefont {L\"{u}}}]{Zhou2022}%
  \BibitemOpen
  \bibfield  {author} {\bibinfo {author} {\bibfnamefont {Y.}~\bibnamefont
  {Zhou}}, \bibinfo {author} {\bibfnamefont {C.-S.}\ \bibnamefont {Hu}},
  \bibinfo {author} {\bibfnamefont {D.-Y.}\ \bibnamefont {L\"{u}}}, \bibinfo
  {author} {\bibfnamefont {X.-K.}\ \bibnamefont {Li}}, \bibinfo {author}
  {\bibfnamefont {H.-M.}\ \bibnamefont {Huang}}, \bibinfo {author}
  {\bibfnamefont {Y.-C.}\ \bibnamefont {Xiong}},\ and\ \bibinfo {author}
  {\bibfnamefont {X.-Y.}\ \bibnamefont {L\"{u}}},\ }\bibfield  {title}
  {\bibinfo {title} {Synergistic enhancement of spin-phonon interaction in a
  hybrid system},\ }\href
  {https://opg.optica.org/prj/abstract.cfm?URI=prj-10-7-1640} {\bibfield
  {journal} {\bibinfo  {journal} {Photon. Res.}\ }\textbf {\bibinfo {volume}
  {10}},\ \bibinfo {pages} {1640} (\bibinfo {year} {2022})}\BibitemShut
  {NoStop}%
\bibitem [{\citenamefont {Zhou}\ \emph {et~al.}(2021)\citenamefont {Zhou},
  \citenamefont {Lü},\ and\ \citenamefont {Zeng}}]{Zhou2021}%
  \BibitemOpen
  \bibfield  {author} {\bibinfo {author} {\bibfnamefont {Y.}~\bibnamefont
  {Zhou}}, \bibinfo {author} {\bibfnamefont {D.-Y.}\ \bibnamefont {Lü}},\ and\
  \bibinfo {author} {\bibfnamefont {W.-Y.}\ \bibnamefont {Zeng}},\ }\bibfield
  {title} {\bibinfo {title} {Chiral single-photon switch-assisted quantum logic
  gate with a nitrogen-vacancy center in a hybrid system},\ }\href
  {https://www.researching.cn/articles/OJ3e767010c3761a11} {\bibfield
  {journal} {\bibinfo  {journal} {Photon. Res.}\ }\textbf {\bibinfo {volume}
  {9}},\ \bibinfo {pages} {405} (\bibinfo {year} {2021})}\BibitemShut {NoStop}%
\bibitem [{\citenamefont {Bennett}\ \emph {et~al.}(2013)\citenamefont
  {Bennett}, \citenamefont {Yao}, \citenamefont {Otterbach}, \citenamefont
  {Zoller}, \citenamefont {Rabl},\ and\ \citenamefont {Lukin}}]{Bennett2013}%
  \BibitemOpen
  \bibfield  {author} {\bibinfo {author} {\bibfnamefont {S.~D.}\ \bibnamefont
  {Bennett}}, \bibinfo {author} {\bibfnamefont {N.~Y.}\ \bibnamefont {Yao}},
  \bibinfo {author} {\bibfnamefont {J.}~\bibnamefont {Otterbach}}, \bibinfo
  {author} {\bibfnamefont {P.}~\bibnamefont {Zoller}}, \bibinfo {author}
  {\bibfnamefont {P.}~\bibnamefont {Rabl}},\ and\ \bibinfo {author}
  {\bibfnamefont {M.~D.}\ \bibnamefont {Lukin}},\ }\bibfield  {title} {\bibinfo
  {title} {Phonon-induced spin-spin interactions in diamond nanostructures:
  Application to spin squeezing},\ }\href
  {https://doi.org/10.1103/PhysRevLett.110.156402} {\bibfield  {journal}
  {\bibinfo  {journal} {Phys. Rev. Lett.}\ }\textbf {\bibinfo {volume} {110}},\
  \bibinfo {pages} {156402} (\bibinfo {year} {2013})}\BibitemShut {NoStop}%
\bibitem [{\citenamefont {Xia}\ and\ \citenamefont {Twamley}(2016)}]{Xia2016}%
  \BibitemOpen
  \bibfield  {author} {\bibinfo {author} {\bibfnamefont {K.}~\bibnamefont
  {Xia}}\ and\ \bibinfo {author} {\bibfnamefont {J.}~\bibnamefont {Twamley}},\
  }\bibfield  {title} {\bibinfo {title} {Generating spin squeezing states and
  greenberger-horne-zeilinger entanglement using a hybrid phonon-spin ensemble
  in diamond},\ }\href {https://doi.org/10.1103/PhysRevB.94.205118} {\bibfield
  {journal} {\bibinfo  {journal} {Phys. Rev. B}\ }\textbf {\bibinfo {volume}
  {94}},\ \bibinfo {pages} {205118} (\bibinfo {year} {2016})}\BibitemShut
  {NoStop}%
\bibitem [{\citenamefont {Ma}\ \emph {et~al.}(2016)\citenamefont {Ma},
  \citenamefont {Zhang}, \citenamefont {Song},\ and\ \citenamefont
  {Wu}}]{MA2016}%
  \BibitemOpen
  \bibfield  {author} {\bibinfo {author} {\bibfnamefont {Y.-H.}\ \bibnamefont
  {Ma}}, \bibinfo {author} {\bibfnamefont {X.-F.}\ \bibnamefont {Zhang}},
  \bibinfo {author} {\bibfnamefont {J.}~\bibnamefont {Song}},\ and\ \bibinfo
  {author} {\bibfnamefont {E.}~\bibnamefont {Wu}},\ }\bibfield  {title}
  {\bibinfo {title} {Bistability and steady-state spin squeezing in diamond
  nanostructures controlled by a nanomechanical resonator},\ }\href
  {https://doi.org/https://doi.org/10.1016/j.aop.2016.03.001} {\bibfield
  {journal} {\bibinfo  {journal} {Annals of Physics}\ }\textbf {\bibinfo
  {volume} {369}},\ \bibinfo {pages} {36} (\bibinfo {year} {2016})}\BibitemShut
  {NoStop}%
\bibitem [{\citenamefont {Li}\ \emph {et~al.}(2020{\natexlab{b}})\citenamefont
  {Li}, \citenamefont {Li}, \citenamefont {Li},\ and\ \citenamefont
  {Li}}]{Li2020}%
  \BibitemOpen
  \bibfield  {author} {\bibinfo {author} {\bibfnamefont {B.}~\bibnamefont
  {Li}}, \bibinfo {author} {\bibfnamefont {X.}~\bibnamefont {Li}}, \bibinfo
  {author} {\bibfnamefont {P.}~\bibnamefont {Li}},\ and\ \bibinfo {author}
  {\bibfnamefont {T.}~\bibnamefont {Li}},\ }\bibfield  {title} {\bibinfo
  {title} {Preparing squeezed spin states in a spin–mechanical hybrid system
  with silicon-vacancy centers},\ }\href
  {https://doi.org/https://doi.org/10.1002/qute.202000034} {\bibfield
  {journal} {\bibinfo  {journal} {Advanced Quantum Technologies}\ }\textbf
  {\bibinfo {volume} {3}},\ \bibinfo {pages} {2000034} (\bibinfo {year}
  {2020}{\natexlab{b}})}\BibitemShut {NoStop}%
\bibitem [{\citenamefont {Chen}\ \emph {et~al.}(2021)\citenamefont {Chen},
  \citenamefont {Qiao}, \citenamefont {Dong}, \citenamefont {Hei},\ and\
  \citenamefont {Li}}]{Chen2021}%
  \BibitemOpen
  \bibfield  {author} {\bibinfo {author} {\bibfnamefont {J.-Q.}\ \bibnamefont
  {Chen}}, \bibinfo {author} {\bibfnamefont {Y.-F.}\ \bibnamefont {Qiao}},
  \bibinfo {author} {\bibfnamefont {X.-L.}\ \bibnamefont {Dong}}, \bibinfo
  {author} {\bibfnamefont {X.-L.}\ \bibnamefont {Hei}},\ and\ \bibinfo {author}
  {\bibfnamefont {P.-B.}\ \bibnamefont {Li}},\ }\bibfield  {title} {\bibinfo
  {title} {Dissipation-assisted preparation of steady spin-squeezed states of
  siv centers},\ }\href {https://doi.org/10.1103/PhysRevA.103.013709}
  {\bibfield  {journal} {\bibinfo  {journal} {Phys. Rev. A}\ }\textbf {\bibinfo
  {volume} {103}},\ \bibinfo {pages} {013709} (\bibinfo {year}
  {2021})}\BibitemShut {NoStop}%
\bibitem [{\citenamefont {Hepp}\ \emph {et~al.}(2014)\citenamefont {Hepp},
  \citenamefont {M\"uller}, \citenamefont {Waselowski}, \citenamefont {Becker},
  \citenamefont {Pingault}, \citenamefont {Sternschulte}, \citenamefont
  {Steinm\"uller-Nethl}, \citenamefont {Gali}, \citenamefont {Maze},
  \citenamefont {Atat\"ure},\ and\ \citenamefont {Becher}}]{Hepp2014}%
  \BibitemOpen
  \bibfield  {author} {\bibinfo {author} {\bibfnamefont {C.}~\bibnamefont
  {Hepp}}, \bibinfo {author} {\bibfnamefont {T.}~\bibnamefont {M\"uller}},
  \bibinfo {author} {\bibfnamefont {V.}~\bibnamefont {Waselowski}}, \bibinfo
  {author} {\bibfnamefont {J.~N.}\ \bibnamefont {Becker}}, \bibinfo {author}
  {\bibfnamefont {B.}~\bibnamefont {Pingault}}, \bibinfo {author}
  {\bibfnamefont {H.}~\bibnamefont {Sternschulte}}, \bibinfo {author}
  {\bibfnamefont {D.}~\bibnamefont {Steinm\"uller-Nethl}}, \bibinfo {author}
  {\bibfnamefont {A.}~\bibnamefont {Gali}}, \bibinfo {author} {\bibfnamefont
  {J.~R.}\ \bibnamefont {Maze}}, \bibinfo {author} {\bibfnamefont
  {M.}~\bibnamefont {Atat\"ure}},\ and\ \bibinfo {author} {\bibfnamefont
  {C.}~\bibnamefont {Becher}},\ }\bibfield  {title} {\bibinfo {title}
  {Electronic structure of the silicon vacancy color center in diamond},\
  }\href {https://doi.org/10.1103/PhysRevLett.112.036405} {\bibfield  {journal}
  {\bibinfo  {journal} {Phys. Rev. Lett.}\ }\textbf {\bibinfo {volume} {112}},\
  \bibinfo {pages} {036405} (\bibinfo {year} {2014})}\BibitemShut {NoStop}%
\bibitem [{\citenamefont {Sipahigil}\ \emph {et~al.}(2016)\citenamefont
  {Sipahigil}, \citenamefont {Evans}, \citenamefont {Sukachev}, \citenamefont
  {Burek}, \citenamefont {Evans}, \citenamefont {Sukachev}, \citenamefont
  {Burek}, \citenamefont {Park},\ and\ \citenamefont {Lukin}}]{Sipahigil2016}%
  \BibitemOpen
  \bibfield  {author} {\bibinfo {author} {\bibfnamefont {A.}~\bibnamefont
  {Sipahigil}}, \bibinfo {author} {\bibfnamefont {R.}~\bibnamefont {Evans}},
  \bibinfo {author} {\bibfnamefont {D.}~\bibnamefont {Sukachev}}, \bibinfo
  {author} {\bibfnamefont {M.}~\bibnamefont {Burek}}, \bibinfo {author}
  {\bibfnamefont {R.}~\bibnamefont {Evans}}, \bibinfo {author} {\bibfnamefont
  {D.}~\bibnamefont {Sukachev}}, \bibinfo {author} {\bibfnamefont
  {M.}~\bibnamefont {Burek}}, \bibinfo {author} {\bibfnamefont
  {H.}~\bibnamefont {Park}},\ and\ \bibinfo {author} {\bibfnamefont
  {M.}~\bibnamefont {Lukin}},\ }\bibfield  {title} {\bibinfo {title} {An
  integrated diamond nanophotonics platform for quantum-optical networks},\
  }\href {https://doi.org/10.1126/science.aah6875} {\bibfield  {journal}
  {\bibinfo  {journal} {Science}\ }\textbf {\bibinfo {volume} {354}},\ \bibinfo
  {pages} {847} (\bibinfo {year} {2016})}\BibitemShut {NoStop}%
\bibitem [{\citenamefont {Zhang}\ \emph
  {et~al.}(2017{\natexlab{b}})\citenamefont {Zhang}, \citenamefont
  {Lagoudakis}, \citenamefont {Tzeng}, \citenamefont {Dory}, \citenamefont
  {Radulaski}, \citenamefont {Kelaita}, \citenamefont {Fischer}, \citenamefont
  {Sun}, \citenamefont {Shen}, \citenamefont {Melosh}, \citenamefont {Chu},\
  and\ \citenamefont {Vu\v{c}kovi\'{c}}}]{Jingyuan2017}%
  \BibitemOpen
  \bibfield  {author} {\bibinfo {author} {\bibfnamefont {J.~L.}\ \bibnamefont
  {Zhang}}, \bibinfo {author} {\bibfnamefont {K.~G.}\ \bibnamefont
  {Lagoudakis}}, \bibinfo {author} {\bibfnamefont {Y.-K.}\ \bibnamefont
  {Tzeng}}, \bibinfo {author} {\bibfnamefont {C.}~\bibnamefont {Dory}},
  \bibinfo {author} {\bibfnamefont {M.}~\bibnamefont {Radulaski}}, \bibinfo
  {author} {\bibfnamefont {Y.}~\bibnamefont {Kelaita}}, \bibinfo {author}
  {\bibfnamefont {K.~A.}\ \bibnamefont {Fischer}}, \bibinfo {author}
  {\bibfnamefont {S.}~\bibnamefont {Sun}}, \bibinfo {author} {\bibfnamefont
  {Z.-X.}\ \bibnamefont {Shen}}, \bibinfo {author} {\bibfnamefont {N.~A.}\
  \bibnamefont {Melosh}}, \bibinfo {author} {\bibfnamefont {S.}~\bibnamefont
  {Chu}},\ and\ \bibinfo {author} {\bibfnamefont {J.}~\bibnamefont
  {Vu\v{c}kovi\'{c}}},\ }\bibfield  {title} {\bibinfo {title} {Complete
  coherent control of silicon vacancies in diamond nanopillars containing
  single defect centers},\ }\href
  {https://opg.optica.org/optica/abstract.cfm?URI=optica-4-11-1317} {\bibfield
  {journal} {\bibinfo  {journal} {Optica}\ }\textbf {\bibinfo {volume} {4}},\
  \bibinfo {pages} {1317} (\bibinfo {year} {2017}{\natexlab{b}})}\BibitemShut
  {NoStop}%
\bibitem [{\citenamefont {Meesala}\ \emph {et~al.}(2018)\citenamefont
  {Meesala}, \citenamefont {Sohn}, \citenamefont {Pingault}, \citenamefont
  {Shao}, \citenamefont {Atikian}, \citenamefont {Holzgrafe}, \citenamefont
  {G\"undo\ifmmode~\breve{g}\else \u{g}\fi{}an}, \citenamefont {Stavrakas},
  \citenamefont {Sipahigil}, \citenamefont {Chia}, \citenamefont {Evans},
  \citenamefont {Burek}, \citenamefont {Zhang}, \citenamefont {Wu},
  \citenamefont {Pacheco}, \citenamefont {Abraham}, \citenamefont {Bielejec},
  \citenamefont {Lukin}, \citenamefont {Atat\"ure},\ and\ \citenamefont
  {Lon\ifmmode~\check{c}\else \v{c}\fi{}ar}}]{Meesala2018}%
  \BibitemOpen
  \bibfield  {author} {\bibinfo {author} {\bibfnamefont {S.}~\bibnamefont
  {Meesala}}, \bibinfo {author} {\bibfnamefont {Y.-I.}\ \bibnamefont {Sohn}},
  \bibinfo {author} {\bibfnamefont {B.}~\bibnamefont {Pingault}}, \bibinfo
  {author} {\bibfnamefont {L.}~\bibnamefont {Shao}}, \bibinfo {author}
  {\bibfnamefont {H.~A.}\ \bibnamefont {Atikian}}, \bibinfo {author}
  {\bibfnamefont {J.}~\bibnamefont {Holzgrafe}}, \bibinfo {author}
  {\bibfnamefont {M.}~\bibnamefont {G\"undo\ifmmode~\breve{g}\else
  \u{g}\fi{}an}}, \bibinfo {author} {\bibfnamefont {C.}~\bibnamefont
  {Stavrakas}}, \bibinfo {author} {\bibfnamefont {A.}~\bibnamefont
  {Sipahigil}}, \bibinfo {author} {\bibfnamefont {C.}~\bibnamefont {Chia}},
  \bibinfo {author} {\bibfnamefont {R.}~\bibnamefont {Evans}}, \bibinfo
  {author} {\bibfnamefont {M.~J.}\ \bibnamefont {Burek}}, \bibinfo {author}
  {\bibfnamefont {M.}~\bibnamefont {Zhang}}, \bibinfo {author} {\bibfnamefont
  {L.}~\bibnamefont {Wu}}, \bibinfo {author} {\bibfnamefont {J.~L.}\
  \bibnamefont {Pacheco}}, \bibinfo {author} {\bibfnamefont {J.}~\bibnamefont
  {Abraham}}, \bibinfo {author} {\bibfnamefont {E.}~\bibnamefont {Bielejec}},
  \bibinfo {author} {\bibfnamefont {M.~D.}\ \bibnamefont {Lukin}}, \bibinfo
  {author} {\bibfnamefont {M.}~\bibnamefont {Atat\"ure}},\ and\ \bibinfo
  {author} {\bibfnamefont {M.}~\bibnamefont {Lon\ifmmode~\check{c}\else
  \v{c}\fi{}ar}},\ }\bibfield  {title} {\bibinfo {title} {Strain engineering of
  the silicon-vacancy center in diamond},\ }\href
  {https://doi.org/10.1103/PhysRevB.97.205444} {\bibfield  {journal} {\bibinfo
  {journal} {Phys. Rev. B}\ }\textbf {\bibinfo {volume} {97}},\ \bibinfo
  {pages} {205444} (\bibinfo {year} {2018})}\BibitemShut {NoStop}%
\bibitem [{\citenamefont {Lemonde}\ \emph {et~al.}(2018)\citenamefont
  {Lemonde}, \citenamefont {Meesala}, \citenamefont {Sipahigil}, \citenamefont
  {Schuetz}, \citenamefont {Lukin}, \citenamefont {Loncar},\ and\ \citenamefont
  {Rabl}}]{Lemonde2018}%
  \BibitemOpen
  \bibfield  {author} {\bibinfo {author} {\bibfnamefont {M.-A.}\ \bibnamefont
  {Lemonde}}, \bibinfo {author} {\bibfnamefont {S.}~\bibnamefont {Meesala}},
  \bibinfo {author} {\bibfnamefont {A.}~\bibnamefont {Sipahigil}}, \bibinfo
  {author} {\bibfnamefont {M.~J.~A.}\ \bibnamefont {Schuetz}}, \bibinfo
  {author} {\bibfnamefont {M.~D.}\ \bibnamefont {Lukin}}, \bibinfo {author}
  {\bibfnamefont {M.}~\bibnamefont {Loncar}},\ and\ \bibinfo {author}
  {\bibfnamefont {P.}~\bibnamefont {Rabl}},\ }\bibfield  {title} {\bibinfo
  {title} {Phonon networks with silicon-vacancy centers in diamond
  waveguides},\ }\href {https://doi.org/10.1103/PhysRevLett.120.213603}
  {\bibfield  {journal} {\bibinfo  {journal} {Phys. Rev. Lett.}\ }\textbf
  {\bibinfo {volume} {120}},\ \bibinfo {pages} {213603} (\bibinfo {year}
  {2018})}\BibitemShut {NoStop}%
\bibitem [{\citenamefont {Qiao}\ \emph {et~al.}(2020)\citenamefont {Qiao},
  \citenamefont {Li}, \citenamefont {Dong}, \citenamefont {Chen}, \citenamefont
  {Zhou},\ and\ \citenamefont {Li}}]{Qiao2020}%
  \BibitemOpen
  \bibfield  {author} {\bibinfo {author} {\bibfnamefont {Y.-F.}\ \bibnamefont
  {Qiao}}, \bibinfo {author} {\bibfnamefont {H.-Z.}\ \bibnamefont {Li}},
  \bibinfo {author} {\bibfnamefont {X.-L.}\ \bibnamefont {Dong}}, \bibinfo
  {author} {\bibfnamefont {J.-Q.}\ \bibnamefont {Chen}}, \bibinfo {author}
  {\bibfnamefont {Y.}~\bibnamefont {Zhou}},\ and\ \bibinfo {author}
  {\bibfnamefont {P.-B.}\ \bibnamefont {Li}},\ }\bibfield  {title} {\bibinfo
  {title} {Phononic-waveguide-assisted steady-state entanglement of
  silicon-vacancy centers},\ }\href
  {https://doi.org/10.1103/PhysRevA.101.042313} {\bibfield  {journal} {\bibinfo
   {journal} {Phys. Rev. A}\ }\textbf {\bibinfo {volume} {101}},\ \bibinfo
  {pages} {042313} (\bibinfo {year} {2020})}\BibitemShut {NoStop}%
\bibitem [{\citenamefont {Ren}\ \emph {et~al.}(2022)\citenamefont {Ren},
  \citenamefont {Feng},\ and\ \citenamefont {Xiang}}]{Ren2022}%
  \BibitemOpen
  \bibfield  {author} {\bibinfo {author} {\bibfnamefont {Z.-Q.}\ \bibnamefont
  {Ren}}, \bibinfo {author} {\bibfnamefont {C.-R.}\ \bibnamefont {Feng}},\ and\
  \bibinfo {author} {\bibfnamefont {Z.-L.}\ \bibnamefont {Xiang}},\ }\bibfield
  {title} {\bibinfo {title} {Deterministic generation of entanglement states
  between silicon-vacancy centers via acoustic modes},\ }\href
  {https://opg.optica.org/oe/abstract.cfm?URI=oe-30-23-41685} {\bibfield
  {journal} {\bibinfo  {journal} {Opt. Express}\ }\textbf {\bibinfo {volume}
  {30}},\ \bibinfo {pages} {41685} (\bibinfo {year} {2022})}\BibitemShut
  {NoStop}%
\bibitem [{\citenamefont {Kepesidis}\ \emph {et~al.}(2016)\citenamefont
  {Kepesidis}, \citenamefont {Lemonde}, \citenamefont {Norambuena},
  \citenamefont {Maze},\ and\ \citenamefont {Rabl}}]{Kepesidis2016}%
  \BibitemOpen
  \bibfield  {author} {\bibinfo {author} {\bibfnamefont {K.~V.}\ \bibnamefont
  {Kepesidis}}, \bibinfo {author} {\bibfnamefont {M.-A.}\ \bibnamefont
  {Lemonde}}, \bibinfo {author} {\bibfnamefont {A.}~\bibnamefont {Norambuena}},
  \bibinfo {author} {\bibfnamefont {J.~R.}\ \bibnamefont {Maze}},\ and\
  \bibinfo {author} {\bibfnamefont {P.}~\bibnamefont {Rabl}},\ }\bibfield
  {title} {\bibinfo {title} {Cooling phonons with phonons: Acoustic reservoir
  engineering with silicon-vacancy centers in diamond},\ }\href
  {https://doi.org/10.1103/PhysRevB.94.214115} {\bibfield  {journal} {\bibinfo
  {journal} {Phys. Rev. B}\ }\textbf {\bibinfo {volume} {94}},\ \bibinfo
  {pages} {214115} (\bibinfo {year} {2016})}\BibitemShut {NoStop}%
\bibitem [{\citenamefont {Sohn}\ \emph {et~al.}(2018)\citenamefont {Sohn},
  \citenamefont {Meesala}, \citenamefont {Pingault}, \citenamefont {Atikian},\
  and\ \citenamefont {Lonar}}]{Sohn2018}%
  \BibitemOpen
  \bibfield  {author} {\bibinfo {author} {\bibfnamefont {Y.~I.}\ \bibnamefont
  {Sohn}}, \bibinfo {author} {\bibfnamefont {S.}~\bibnamefont {Meesala}},
  \bibinfo {author} {\bibfnamefont {B.}~\bibnamefont {Pingault}}, \bibinfo
  {author} {\bibfnamefont {H.~A.}\ \bibnamefont {Atikian}},\ and\ \bibinfo
  {author} {\bibfnamefont {M.}~\bibnamefont {Lonar}},\ }\bibfield  {title}
  {\bibinfo {title} {Controlling the coherence of a diamond spin qubit through
  its strain environment},\ }\href
  {https://www.nature.com/articles/s41467-018-04340-3} {\bibfield  {journal}
  {\bibinfo  {journal} {Nat. Commun.}\ }\textbf {\bibinfo {volume} {9}},\
  \bibinfo {pages} {2012} (\bibinfo {year} {2018})}\BibitemShut {NoStop}%
\bibitem [{\citenamefont {Dong}\ \emph {et~al.}(2021)\citenamefont {Dong},
  \citenamefont {Li}, \citenamefont {Liu},\ and\ \citenamefont
  {Nori}}]{Dong2021}%
  \BibitemOpen
  \bibfield  {author} {\bibinfo {author} {\bibfnamefont {X.-L.}\ \bibnamefont
  {Dong}}, \bibinfo {author} {\bibfnamefont {P.-B.}\ \bibnamefont {Li}},
  \bibinfo {author} {\bibfnamefont {T.}~\bibnamefont {Liu}},\ and\ \bibinfo
  {author} {\bibfnamefont {F.}~\bibnamefont {Nori}},\ }\bibfield  {title}
  {\bibinfo {title} {Unconventional quantum sound-matter interactions in
  spin-optomechanical-crystal hybrid systems},\ }\href
  {https://doi.org/10.1103/PhysRevLett.126.203601} {\bibfield  {journal}
  {\bibinfo  {journal} {Phys. Rev. Lett.}\ }\textbf {\bibinfo {volume} {126}},\
  \bibinfo {pages} {203601} (\bibinfo {year} {2021})}\BibitemShut {NoStop}%
\bibitem [{\citenamefont {Fr\"ohlich}(1950)}]{Fr1950}%
  \BibitemOpen
  \bibfield  {author} {\bibinfo {author} {\bibfnamefont {H.}~\bibnamefont
  {Fr\"ohlich}},\ }\bibfield  {title} {\bibinfo {title} {Theory of the
  superconducting state. i. the ground state at the absolute zero of
  temperature},\ }\href {https://doi.org/10.1103/PhysRev.79.845} {\bibfield
  {journal} {\bibinfo  {journal} {Phys. Rev.}\ }\textbf {\bibinfo {volume}
  {79}},\ \bibinfo {pages} {845} (\bibinfo {year} {1950})}\BibitemShut
  {NoStop}%
\bibitem [{\citenamefont {Nakajima}(1955)}]{Nakajima1955}%
  \BibitemOpen
  \bibfield  {author} {\bibinfo {author} {\bibfnamefont {S.}~\bibnamefont
  {Nakajima}},\ }\bibfield  {title} {\bibinfo {title} {Perturbation theory in
  statistical mechanics},\ }\href {https://doi.org/10.1080/00018735500101254}
  {\bibfield  {journal} {\bibinfo  {journal} {Advances in Physics}\ }\textbf
  {\bibinfo {volume} {4}},\ \bibinfo {pages} {363} (\bibinfo {year}
  {1955})}\BibitemShut {NoStop}%
\bibitem [{\citenamefont {Schrieffer}\ and\ \citenamefont
  {Wolff}(1966)}]{Schrieffer1966}%
  \BibitemOpen
  \bibfield  {author} {\bibinfo {author} {\bibfnamefont {J.~R.}\ \bibnamefont
  {Schrieffer}}\ and\ \bibinfo {author} {\bibfnamefont {P.~A.}\ \bibnamefont
  {Wolff}},\ }\bibfield  {title} {\bibinfo {title} {Relation between the
  anderson and kondo hamiltonians},\ }\href
  {https://doi.org/10.1103/PhysRev.149.491} {\bibfield  {journal} {\bibinfo
  {journal} {Phys. Rev.}\ }\textbf {\bibinfo {volume} {149}},\ \bibinfo {pages}
  {491} (\bibinfo {year} {1966})}\BibitemShut {NoStop}%
\bibitem [{\citenamefont {Bravyi}\ \emph {et~al.}(2011)\citenamefont {Bravyi},
  \citenamefont {DiVincenzo},\ and\ \citenamefont {Loss}}]{Sergey2011}%
  \BibitemOpen
  \bibfield  {author} {\bibinfo {author} {\bibfnamefont {S.}~\bibnamefont
  {Bravyi}}, \bibinfo {author} {\bibfnamefont {D.~P.}\ \bibnamefont
  {DiVincenzo}},\ and\ \bibinfo {author} {\bibfnamefont {D.}~\bibnamefont
  {Loss}},\ }\bibfield  {title} {\bibinfo {title} {Schrieffer–wolff
  transformation for quantum many-body systems},\ }\href
  {https://doi.org/https://doi.org/10.1016/j.aop.2011.06.004} {\bibfield
  {journal} {\bibinfo  {journal} {Annals of Physics}\ }\textbf {\bibinfo
  {volume} {326}},\ \bibinfo {pages} {2793} (\bibinfo {year}
  {2011})}\BibitemShut {NoStop}%
\bibitem [{\citenamefont {Blais}\ \emph {et~al.}(2007)\citenamefont {Blais},
  \citenamefont {Gambetta}, \citenamefont {Wallraff}, \citenamefont {Schuster},
  \citenamefont {Girvin}, \citenamefont {Devoret},\ and\ \citenamefont
  {Schoelkopf}}]{Blais2007}%
  \BibitemOpen
  \bibfield  {author} {\bibinfo {author} {\bibfnamefont {A.}~\bibnamefont
  {Blais}}, \bibinfo {author} {\bibfnamefont {J.}~\bibnamefont {Gambetta}},
  \bibinfo {author} {\bibfnamefont {A.}~\bibnamefont {Wallraff}}, \bibinfo
  {author} {\bibfnamefont {D.~I.}\ \bibnamefont {Schuster}}, \bibinfo {author}
  {\bibfnamefont {S.~M.}\ \bibnamefont {Girvin}}, \bibinfo {author}
  {\bibfnamefont {M.~H.}\ \bibnamefont {Devoret}},\ and\ \bibinfo {author}
  {\bibfnamefont {R.~J.}\ \bibnamefont {Schoelkopf}},\ }\bibfield  {title}
  {\bibinfo {title} {Quantum-information processing with circuit quantum
  electrodynamics},\ }\href {https://doi.org/10.1103/PhysRevA.75.032329}
  {\bibfield  {journal} {\bibinfo  {journal} {Phys. Rev. A}\ }\textbf {\bibinfo
  {volume} {75}},\ \bibinfo {pages} {032329} (\bibinfo {year}
  {2007})}\BibitemShut {NoStop}%
\bibitem [{\citenamefont {Wineland}\ \emph {et~al.}(1992)\citenamefont
  {Wineland}, \citenamefont {Bollinger}, \citenamefont {Itano}, \citenamefont
  {Moore},\ and\ \citenamefont {Heinzen}}]{Wineland1992}%
  \BibitemOpen
  \bibfield  {author} {\bibinfo {author} {\bibfnamefont {D.~J.}\ \bibnamefont
  {Wineland}}, \bibinfo {author} {\bibfnamefont {J.~J.}\ \bibnamefont
  {Bollinger}}, \bibinfo {author} {\bibfnamefont {W.~M.}\ \bibnamefont
  {Itano}}, \bibinfo {author} {\bibfnamefont {F.~L.}\ \bibnamefont {Moore}},\
  and\ \bibinfo {author} {\bibfnamefont {D.~J.}\ \bibnamefont {Heinzen}},\
  }\bibfield  {title} {\bibinfo {title} {Spin squeezing and reduced quantum
  noise in spectroscopy},\ }\href {https://doi.org/10.1103/PhysRevA.46.R6797}
  {\bibfield  {journal} {\bibinfo  {journal} {Phys. Rev. A}\ }\textbf {\bibinfo
  {volume} {46}},\ \bibinfo {pages} {R6797} (\bibinfo {year}
  {1992})}\BibitemShut {NoStop}%
\bibitem [{\citenamefont {Wang}\ \emph
  {et~al.}(2017{\natexlab{b}})\citenamefont {Wang}, \citenamefont {Zhou},
  \citenamefont {Zhang}, \citenamefont {Liu}, \citenamefont {Li}, \citenamefont
  {Li}, \citenamefont {Liu}, \citenamefont {Wang},\ and\ \citenamefont
  {Gao}}]{WangJunfeng2017}%
  \BibitemOpen
  \bibfield  {author} {\bibinfo {author} {\bibfnamefont {J.}~\bibnamefont
  {Wang}}, \bibinfo {author} {\bibfnamefont {Y.}~\bibnamefont {Zhou}}, \bibinfo
  {author} {\bibfnamefont {X.}~\bibnamefont {Zhang}}, \bibinfo {author}
  {\bibfnamefont {F.}~\bibnamefont {Liu}}, \bibinfo {author} {\bibfnamefont
  {Y.}~\bibnamefont {Li}}, \bibinfo {author} {\bibfnamefont {K.}~\bibnamefont
  {Li}}, \bibinfo {author} {\bibfnamefont {Z.}~\bibnamefont {Liu}}, \bibinfo
  {author} {\bibfnamefont {G.}~\bibnamefont {Wang}},\ and\ \bibinfo {author}
  {\bibfnamefont {W.}~\bibnamefont {Gao}},\ }\bibfield  {title} {\bibinfo
  {title} {Efficient generation of an array of single silicon-vacancy defects
  in silicon carbide},\ }\href
  {https://doi.org/10.1103/PhysRevApplied.7.064021} {\bibfield  {journal}
  {\bibinfo  {journal} {Phys. Rev. Applied}\ }\textbf {\bibinfo {volume} {7}},\
  \bibinfo {pages} {064021} (\bibinfo {year} {2017}{\natexlab{b}})}\BibitemShut
  {NoStop}%
\bibitem [{\citenamefont {J.~Pla}\ \emph {et~al.}(2012)\citenamefont {J.~Pla},
  \citenamefont {Tan}, \citenamefont {P.~Dehollain}, \citenamefont {H.~Lim},
  \citenamefont {J.~L.~Morton}, \citenamefont {N.~Jamieson}, \citenamefont
  {S.~Dzurak},\ and\ \citenamefont {Morello}}]{Jarryd2012}%
  \BibitemOpen
  \bibfield  {author} {\bibinfo {author} {\bibfnamefont {J.}~\bibnamefont
  {J.~Pla}}, \bibinfo {author} {\bibfnamefont {K.~Y.}\ \bibnamefont {Tan}},
  \bibinfo {author} {\bibfnamefont {J.}~\bibnamefont {P.~Dehollain}}, \bibinfo
  {author} {\bibfnamefont {W.}~\bibnamefont {H.~Lim}}, \bibinfo {author}
  {\bibfnamefont {J.}~\bibnamefont {J.~L.~Morton}}, \bibinfo {author}
  {\bibfnamefont {D.}~\bibnamefont {N.~Jamieson}}, \bibinfo {author}
  {\bibfnamefont {A.}~\bibnamefont {S.~Dzurak}},\ and\ \bibinfo {author}
  {\bibfnamefont {A.}~\bibnamefont {Morello}},\ }\bibfield  {title} {\bibinfo
  {title} {A single-atom electron spin qubit in silicon},\ }\href
  {https://www.nature.com/articles/nature11449} {\bibfield  {journal} {\bibinfo
   {journal} {Nature}\ }\textbf {\bibinfo {volume} {489}},\ \bibinfo {pages}
  {541} (\bibinfo {year} {2012})}\BibitemShut {NoStop}%
\bibitem [{\citenamefont {Pingault}\ \emph {et~al.}(2014)\citenamefont
  {Pingault}, \citenamefont {Becker}, \citenamefont {Schulte}, \citenamefont
  {Arend}, \citenamefont {Hepp}, \citenamefont {Godde}, \citenamefont
  {Tartakovskii}, \citenamefont {Markham}, \citenamefont {Becher},\ and\
  \citenamefont {Atat\"ure}}]{Pingault2014}%
  \BibitemOpen
  \bibfield  {author} {\bibinfo {author} {\bibfnamefont {B.}~\bibnamefont
  {Pingault}}, \bibinfo {author} {\bibfnamefont {J.~N.}\ \bibnamefont
  {Becker}}, \bibinfo {author} {\bibfnamefont {C.~H.~H.}\ \bibnamefont
  {Schulte}}, \bibinfo {author} {\bibfnamefont {C.}~\bibnamefont {Arend}},
  \bibinfo {author} {\bibfnamefont {C.}~\bibnamefont {Hepp}}, \bibinfo {author}
  {\bibfnamefont {T.}~\bibnamefont {Godde}}, \bibinfo {author} {\bibfnamefont
  {A.~I.}\ \bibnamefont {Tartakovskii}}, \bibinfo {author} {\bibfnamefont
  {M.}~\bibnamefont {Markham}}, \bibinfo {author} {\bibfnamefont
  {C.}~\bibnamefont {Becher}},\ and\ \bibinfo {author} {\bibfnamefont
  {M.}~\bibnamefont {Atat\"ure}},\ }\bibfield  {title} {\bibinfo {title}
  {All-optical formation of coherent dark states of silicon-vacancy spins in
  diamond},\ }\href {https://doi.org/10.1103/PhysRevLett.113.263601} {\bibfield
   {journal} {\bibinfo  {journal} {Phys. Rev. Lett.}\ }\textbf {\bibinfo
  {volume} {113}},\ \bibinfo {pages} {263601} (\bibinfo {year}
  {2014})}\BibitemShut {NoStop}%
\bibitem [{\citenamefont {Sukachev}\ \emph {et~al.}(2017)\citenamefont
  {Sukachev}, \citenamefont {Sipahigil}, \citenamefont {Nguyen}, \citenamefont
  {Bhaskar}, \citenamefont {Evans}, \citenamefont {Jelezko},\ and\
  \citenamefont {Lukin}}]{Sukachev2017}%
  \BibitemOpen
  \bibfield  {author} {\bibinfo {author} {\bibfnamefont {D.~D.}\ \bibnamefont
  {Sukachev}}, \bibinfo {author} {\bibfnamefont {A.}~\bibnamefont {Sipahigil}},
  \bibinfo {author} {\bibfnamefont {C.~T.}\ \bibnamefont {Nguyen}}, \bibinfo
  {author} {\bibfnamefont {M.~K.}\ \bibnamefont {Bhaskar}}, \bibinfo {author}
  {\bibfnamefont {R.~E.}\ \bibnamefont {Evans}}, \bibinfo {author}
  {\bibfnamefont {F.}~\bibnamefont {Jelezko}},\ and\ \bibinfo {author}
  {\bibfnamefont {M.~D.}\ \bibnamefont {Lukin}},\ }\bibfield  {title} {\bibinfo
  {title} {Silicon-vacancy spin qubit in diamond: A quantum memory exceeding 10
  ms with single-shot state readout},\ }\href
  {https://doi.org/10.1103/PhysRevLett.119.223602} {\bibfield  {journal}
  {\bibinfo  {journal} {Phys. Rev. Lett.}\ }\textbf {\bibinfo {volume} {119}},\
  \bibinfo {pages} {223602} (\bibinfo {year} {2017})}\BibitemShut {NoStop}%
\bibitem [{\citenamefont {Khanaliloo}\ \emph {et~al.}(2015)\citenamefont
  {Khanaliloo}, \citenamefont {Jayakumar}, \citenamefont {Hryciw},
  \citenamefont {Lake}, \citenamefont {Kaviani},\ and\ \citenamefont
  {Barclay}}]{Khanaliloo2015}%
  \BibitemOpen
  \bibfield  {author} {\bibinfo {author} {\bibfnamefont {B.}~\bibnamefont
  {Khanaliloo}}, \bibinfo {author} {\bibfnamefont {H.}~\bibnamefont
  {Jayakumar}}, \bibinfo {author} {\bibfnamefont {A.~C.}\ \bibnamefont
  {Hryciw}}, \bibinfo {author} {\bibfnamefont {D.~P.}\ \bibnamefont {Lake}},
  \bibinfo {author} {\bibfnamefont {H.}~\bibnamefont {Kaviani}},\ and\ \bibinfo
  {author} {\bibfnamefont {P.~E.}\ \bibnamefont {Barclay}},\ }\bibfield
  {title} {\bibinfo {title} {Single-crystal diamond nanobeam waveguide
  optomechanics},\ }\href {https://doi.org/10.1103/PhysRevX.5.041051}
  {\bibfield  {journal} {\bibinfo  {journal} {Phys. Rev. X}\ }\textbf {\bibinfo
  {volume} {5}},\ \bibinfo {pages} {041051} (\bibinfo {year}
  {2015})}\BibitemShut {NoStop}%
\bibitem [{\citenamefont {Ovartchaiyapong}\ \emph {et~al.}(2012)\citenamefont
  {Ovartchaiyapong}, \citenamefont {Pascal}, \citenamefont {Myers},
  \citenamefont {Lauria},\ and\ \citenamefont
  {Bleszynski~Jayich}}]{Ovartchaiyapong2012}%
  \BibitemOpen
  \bibfield  {author} {\bibinfo {author} {\bibfnamefont {P.}~\bibnamefont
  {Ovartchaiyapong}}, \bibinfo {author} {\bibfnamefont {L.~M.~A.}\ \bibnamefont
  {Pascal}}, \bibinfo {author} {\bibfnamefont {B.~A.}\ \bibnamefont {Myers}},
  \bibinfo {author} {\bibfnamefont {P.}~\bibnamefont {Lauria}},\ and\ \bibinfo
  {author} {\bibfnamefont {A.~C.}\ \bibnamefont {Bleszynski~Jayich}},\
  }\bibfield  {title} {\bibinfo {title} {High quality factor single-crystal
  diamond mechanical resonators},\ }\href {https://doi.org/10.1063/1.4760274}
  {\bibfield  {journal} {\bibinfo  {journal} {Appl. Phys. Lett.}\ }\textbf
  {\bibinfo {volume} {101}},\ \bibinfo {pages} {163505} (\bibinfo {year}
  {2012})}\BibitemShut {NoStop}%
\end{thebibliography}%

\end{document}